\documentclass[lettersize,journal]{IEEEtran}
\usepackage{amsmath,amsfonts}
\usepackage{algorithmic}
\usepackage{algorithm}
\usepackage{array}
\usepackage[caption=false,font=normalsize,labelfont=sf,textfont=sf]{subfig}
\usepackage{textcomp}
\usepackage{stfloats}
\usepackage{url}
\usepackage{verbatim}
\usepackage{graphicx}
\usepackage{cite}
\begin{document}


\title{A DRL-Driven Optimization of RAN Slice Resource Partitioning for V2X SLA Compliance in 5G Networks}

\author{M. Martínez, I. de-la-Bandera, D. E. García, P. Vera, S. Fortes,\textit{ Senior Member, IEEE}, M. L. Luque, A. Mendo, J. Ramiro, and R. Barco {}
\thanks{This work has been submitted to the IEEE for possible publication. Copyright may be transferred without notice, after which this
version may no longer be accessible. This work has been partially funded by Ministerio para la Transformaci\'on Digital y de la Función Pública and European Union - NextGenerationEU within the framework ``Recuperaci\'on, Transformaci\'on y Resiliencia'' under the project 5GVEC.}
\thanks{M. Martínez, I. de-la-Bandera, D. E. García, P. Vera, S. Fortes, and R. Barco are with the Telecommunications Research Institute
(TELMA), E.T.S.I de Telecomunicación, University of Málaga, 29010 Málaga,
Spain (e-mail: mcruces@uma.es; ibanderac@uma.es; danielgarciaf@uma.es; pablovera@uma.es; sergio.fortes@uma.es; rbarco@uma.es).}
\thanks{ M. L. Luque, A. Mendo, and J. Ramiro are with Ericsson, C Severo Ochoa 55, 29590 Málaga, Spain (e-mail: maria.laura.luque@ericsson.com, adriano.mendo@ericsson.com, juan.ramiro@ericsson.com).}
}


\makeatletter
\def\ps@myIEEEstyle{%
  \def\@oddhead{%
    \scriptsize
    This work has been submitted to the IEEE for possible publication.
    Copyright may be transferred without notice, after which this version
    may no longer be accessible.
    \hfill\thepage
  }%
  \def\@evenhead{%
    \scriptsize
    This work has been submitted to the IEEE for possible publication.
    Copyright may be transferred without notice, after which this version
    may no longer be accessible.
    \hfill\thepage
  }%
  \def\@oddfoot{}%
  \def\@evenfoot{}%
}
\makeatother

\maketitle

\pagestyle{myIEEEstyle}
\thispagestyle{myIEEEstyle}

\begin{abstract}
Vehicle-to-Everything (V2X) communications impose very demanding requirements in terms of latency and reliability, which must be met in scenarios where multiple services with diverse performance targets coexist. In such scenarios, traffic-intensive services compete for limited radio resources, complicating the fulfillment of V2X service demands. Within this context, Network Slicing (NS) emerges as a key factor that enables the creation of multiple slices and the allocation of resources among them to satisfy heterogeneous service requirements. In particular, this work addresses the Radio Access Network (RAN) slicing problem from the perspective of Physical Resource Block (PRB) partitioning under high traffic demand conditions. To this end, a reinforcement learning approach based on Proximal Policy Optimization (PPO) is proposed to determine PRB allocations that satisfy the strict latency and reliability requirements of V2X services, while improving resource utilization efficiency and minimizing performance degradation of enhanced Mobile BroadBand (eMBB) services. The proposed solution is evaluated through simulation-based experiments under various traffic loads and different V2X service requirements, demonstrating its ability to adapt resource partitioning to network conditions and service demands. 
\end{abstract}


\section{Introduction}
\IEEEPARstart{T}{he} development of the Fifth-Generation (5G) mobile networks has significantly improved data rate, latency, reliability, and connectivity, enabling cellular systems to support services with diverse and stringent performance requirements. Among these, Vehicle-to-Everything (V2X) applications demand extremely low latency and high reliability. 

In Long-Term Evolution (LTE) systems, such requirements were achieved through Vehicle-to-Vehicle (V2V) and Vehicle-to-Infrastructure (V2I) communications over the PC5 interface, whereas the Uu interface was mainly reserved for infotainment and other non-critical applications \cite{cita_pc5_uu}. However, as highlighted in \cite{cita_pc5_uu}, the introduction of 5G New Radio (NR) and its improved latency and reliability capabilities enable the Uu interface to also support advanced V2X services.

Nevertheless, cellular network traffic originates from different service types with diverse performance targets, including enhanced Mobile BroadBand (eMBB), Ultra-Reliable Low-Latency Communications (URLLC), and massive Machine-Type Communications (mMTC). In this context, supporting V2X applications poses significant challenges for network resource management, as their Service Level Agreements (SLA) must be satisfied while minimizing the impact on the performance of other coexisting services. This challenge is exacerbated under high traffic conditions and in the presence of bandwidth-intensive services. Moreover, V2X use cases themselves exhibit heterogeneous requirements, further increasing the complexity of resource allocation \cite{cita_multiagent}. 

Together, these observations reveal the need for a flexible and adaptive network architecture capable of accommodating services with heterogeneous performance targets. To achieve such flexibility, Network Slicing (NS) emerges as a key enabler in 5G networks, as it allows the creation of multiple logical networks, or network slices, over a shared physical infrastructure \cite{cita_an_efficent}. As described in \cite{cita_an_efficent}, each segment can be tailored to the specific requirements of a service or group of services to achieve optimal performance in terms of latency, reliability, and throughput. To this end, resources across the network must be appropriately distributed across the slices.

Although slicing can be applied both at the Core network and Radio Access Network (RAN) levels, RAN slicing has been identified as one of the most promising techniques for delivering customized services, including V2X communications \cite{cita_aided_decoupled}, \cite{cita_rl_dynamic}. As discussed in \cite{cita_an_efficent}, efficient RAN slicing can improve network capacity utilization, reduce traffic congestion, and ensure high Quality of Service (QoS) in terms of latency, which is crucial for V2X applications, but particularly challenging due to the limited radio resources and the dynamic nature of wireless channels \cite{cita_multiagent}. 
\IEEEpubidadjcol
These problems are further amplified in V2X communications, where high vehicle mobility, rapidly changing network topologies, and time-varying SLAs make resource allocation particularly critical \cite{cita_multiagent}. Furthermore, the variation in the number of vehicles and in demand over time requires adjustments to resource allocation, since an inadequate one can lead to over-provisioning or under-provisioning, which risks compliance with the SLAs and degrades performance within the network.

These considerations underscore the need for intelligent and adaptive RAN resource management strategies that adjust radio resources to network conditions and the needs of V2X services. This has motivated the application of AI algorithms, which have the potential to handle the complexity and dynamism of wireless environments more efficiently than conventional optimization methods \cite{cita_genAI}. According to \cite{cita_genAI}, the integration of such tools into RAN management can make a significant difference in tasks, such as physical resource block (PRB) allocation and power control.

Motivated by these challenges and opportunities, this work addresses the problem of radio resource management in downlink (DL) by optimizing the partitioning of radio resources among network slices through a Proximal Policy Optimization (PPO)-based Reinforcement Learning (RL) algorithm to enable ultra-reliable and low latency V2X applications over the Uu interface. Among the available radio resources, the optimization is focused on PRBs, which are the basic time-frequency allocation units in wireless communication systems. Specifically, the goal is to determine the optimal PRB allocation that guarantees low latency and high reliability for V2X traffic, and that maximizes overall resource efficiency while minimizing the impact on eMBB users.


\section{Related Work}

Traditionally, resource allocation in NS has been addressed through mathematical optimization. These methods have been applied in works such as \cite{cita_trad1}, which jointly optimizes bandwidth allocation and power consumption while satisfying URLLC latency constraints and the quality of eMBB services, and in \cite{cita_trad2}, where network and edge computation resources are sliced to minimize end-to-end latency under specific constraints. Heuristic and meta-heuristic approaches have also been explored as an alternative to AI-based solutions. For instance, \cite{cita_trad3} proposes a rule-based heuristic approach named Adaptive Hungarian Algorithm (AHA) to solve the network slicing allocation problem, whereas authors in \cite{cita_trad4} address the RAN slicing problem using genetic algorithms to allocate Resource Blocks (RBs) to each slice considering the users' QoS requirements. 

Nevertheless, mathematical optimization methods typically rely on static models that lack adaptability to the changing network conditions, and although heuristic and meta-heuristic methods can find good solutions rapidly, their performance may degrade when network conditions deviate from the design assumptions \cite{cita_genAI}. Consequently, data-driven approaches have attracted increasing attention for dynamic slice management.

Among them, \cite{carlos} proposes a framework to support network slice negotiation and maintenance through the estimation of Key Quality Indicators (KQIs) using regression models. These KQIs estimations and dynamic thresholds are exploited to allocate resources according to the radio conditions to satisfy the service requirements. 
Recently, RL and Deep Reinforcement Learning (DRL) have emerged as a promising solution for the RAN slicing problem. In \cite{cita_an_efficent}, an offline Q-Learning approach is proposed to determine the slice ratios that maximize resource utilization while ensuring an outage probability lower than a maximum tolerable limit in an eMBB-V2X scenario. 
Other works, such as \cite{dynamic} and \cite{cita_rl_dynamic}, consider slicing environments with heterogeneous traffic types, including URLLC. However, instead of addressing reliability and low latency as QoS constraints, the former allocates bandwidth among slices with the goal of minimizing the blocked requests in the system using a Deep Q-network (DQN), while the latter defines the reward of a Q-Learning approach as the priority-weighted sum of requests satisfied.

It is noteworthy that, unlike the present contribution, most of these frameworks focus on generic slice profiles or do not explicitly consider the stringent latency and reliability requirements from V2X services. 

Other works go further by addressing performance optimization from the perspective of latency. For example, in \cite{real}, the slicing optimization problem incorporates latency compliance for a URLLC slice modeled as an autonomous driving system by penalizing whenever the average delay exceeds a predefined threshold. Meanwhile, \cite{comparative_analysis} employs a PPO-based agent to reconfigure Base Station control parameters to maximize eMBB user throughput and the number of transmitted mMTC packets while minimizing the URLLC packet latency by accounting for the buffer occupancy of the URLLC slice. 
In the same way, \cite{TNSM_1} proposes a Double Deep Q-Network (DDQN)-based approach to minimize system costs while satisfying QoS constraints. These QoS constraints are defined according to the slice type; eMBB users must satisfy a minimum data rate requirement, URLLC users a maximum packet delay requirement, and mMTC users a minimum Signal-to-Noise Ratio (SNR) requirement. Yet, reliability requirements are not explicitly addressed in either of these works, and the scenarios include only a URLLC, an eMBB, and an mMTC service, thus modeling only inter-service heterogeneity. As a result, competition for resources between V2X services with different traffic patterns and demands, which increases the dimensionality of the problem and the complexity of decision-making, is not considered. 




In contrast, \cite{DRL} proposes a DRL-based solution that seeks slicing policies that satisfy both latency and reliability constraints, while \cite{DDPG} employs a Deterministic Policy Gradient (DDPG) algorithm to perform the inter-slice resource allocation to meet the QoS of specific vehicular applications. Although both demonstrate adaptability to varying SLAs and vehicular densities, respectively, the evaluation is limited to only two V2X slices. Consequently, the complexity associated with dimensioning low-latency and highly reliable services in scenarios with services whose performance targets are defined by different metrics, e.g., throughput, is not fully addressed. Furthermore, in \cite{DDPG}, reliability is captured through the Packet Delivery Ratio (PDR), and latency requirements are not addressed at this decision level. Finally, \cite{cita_multiagent} proposes a two-timescale resource allocation framework where a PPO-based multi-agent DRL (MADRL) algorithm determines the bandwidth assigned to the different slices. The framework considers high-capacity V2I services and V2V services with stringent latency and reliability requirements. However, these requirements are modeled through probabilistic constraints on the queue length and the received SNR, whereas our work defines V2X SLA compliance directly in terms of packet delay that jointly captures latency and reliability and prioritizes its satisfaction over throughput optimization.

In addition to the optimization objectives, various works have compared different RL algorithms, demonstrating how PPO outperforms other approaches like DQN and Q-Learning. In \cite{DRL}, the PPO-based approach achieves higher episode rewards and lower SLA violation rates while adapting to changing conditions. Similarly, \cite{real} demonstrates how PPO achieves a more robust and consistent performance across the scenarios and achieves higher throughput than DQN, while achieving a lower latency for a greater percentage of packets. These results highlight the suitability of PPO for dynamic radio resource allocation problems and motivated its adoption in this work. Since PPO implements a stable policy optimization mechanism, excessively large policy updates are avoided, improving training stability and enabling efficient learning in dynamic environments.

The contributions of this paper are summarized as follows:
\begin{itemize}
    \item A DRL algorithm based on PPO for RAN slicing in heterogeneous multi-service scenarios where multiple V2X and eMBB slices with different traffic characteristics and performance targets coexist, which increases the problem's dimensionality and the complexity of the decision-making process. 
    
    
    \item A reward function that enables the joint optimization of V2X SLA compliance, radio resource utilization, and eMBB throughput while ensuring balance between throughput-demanding services. Unlike most existing works, V2X SLA compliance is defined by jointly considering both delay and reliability.
    
    
    
    \item A simulation-based evaluation considering different performance and network load conditions, as well as various V2X traffic patterns, both adapted from 3rd Generation Partnership Project (3GPP) to demonstrate the adaptability of the approach.
\end{itemize}

\section{System Model and Problem Formulation}
\label{sec:sys_model}
\subsection{System Model}

The considered scenario is limited to the DL of a 5G network, where the gNodeB (gNB) serves a set $U$ of mobile users distributed within the coverage area of a cell of interest. The cell radio capacity is expressed in terms of PRBs, with $C_{\mathrm{tot}}$ denoting the total number of PRBs available in a cell. Resource allocation decisions are performed at the cell level, and PRBs are assigned by the scheduler on a per-slot basis, assuming a Transmission Time Interval (TTI) of 1ms.

The network supports RAN slicing by deploying a set of slices $N$, with each slice $n_{i}$ associated with a specific service.

Even though two main service types are considered, V2X and eMBB, each of them is further differentiated into multiple services with distinct traffic characteristics and performance requirements. This approach allows for the analysis of heterogeneous vehicular applications, as well as eMBB services with different achievable data rates.

\subsubsection{V2X Service Model}
The V2X services considered in this work generate traffic patterns modeled as periodic packet transmissions compatible with the 3GPP definition for V2X communications. Accordingly, for a V2X service $i$, a packet of $s^{\mathrm{v2x}}_{i}$ bits is generated every $T^{\mathrm{v2x}}_{i}$ milliseconds for each user. This results in the per-user packet arrival rate given by (\ref{lambdav2x}), where $\lambda^{\mathrm{v2x}}_{i,u}$ is expressed in packets per millisecond and $u$ refers to a specific user associated with the V2X service $i$.
\begin{equation}
\label{lambdav2x}
\lambda^{\mathrm{v2x}}_{i,u} = \frac{1}{T^{\mathrm{v2x}}_{i}}
\end{equation}


At the service level, the packet arrival rate is characterized by the average number of packets that arrive at the DL buffer per millisecond. Being $A^{\mathrm{v2x}}_i(t)$ the packets generated for V2X service $i$ at a TTI $t$, $\mathcal{T}$ the set of TTIs within the observation interval, $|\mathcal{T}|$ its cardinality, and $\Delta{t}$ the TTI duration (1ms), the service-level packet arrival rate is defined as:

\begin{equation}
\label{lambda_service}
\lambda^{\mathrm{v2x}}_i
=
\frac{1}{|\mathcal{T}|\Delta{t}}
\sum_{t \in \mathcal{T}} A^{\mathrm{v2x}}_i(t).
\end{equation}

To characterize the performance of V2X services, the SLA of a V2X service is defined as the tuple $\mathrm{SLA}^{\mathrm{v2x}}_{i}=(D^{\mathrm{RAN}}_{\mathrm{max}, i}, \varGamma_i)$, where $D^{\mathrm{RAN}}_{\mathrm{max},i}$ is the maximum tolerable delay experienced by a packet in the DL and $\varGamma_i$ the required reliability percentage for the service $i$. 

The assessment of compliance with the different SLA targets can be performed jointly by analyzing the percentiles of the delay distributions. This approach aligns with the definition of reliability presented in \cite{mlucas}, where it is understood as the percentage of packets that must satisfy a given latency constraint.


Being $D^{\mathrm{v2x}}_{i}$ the set of 
delay samples collected for the service $i$ during $\mathcal{T}$, the service performance can be characterized by the $\varGamma$-th percentile of the measured 
experienced packet delays, which is defined as $P_{\varGamma, i}=\mathrm{percentile}_{\varGamma}(D^{\mathrm{v2x}}_{i})$. Consequently, the SLA of the V2X service is considered satisfied if $P_{\varGamma, i} \leq D^{\mathrm{RAN}}_{\mathrm{max}}$.

According to the 3GPP specification TS 23.501 \cite{5QI_table}, the Packet Delay Budget (PDB) defines an upper bound for the time that a packet may be delayed between the User Equipment (UE) and the N6 termination point at the User Plane Function (UPF). In the case of (Guaranteed bit Rate) GBR QoS Flows using the Delay-critical GBR resource type, packets whose delay exceeds the PDB are considered lost and can be discarded or delivered depending on local decision. 

In this work, these packets will be discarded, since V2X applications are delay-critical and packets that arrive after the deadline may carry outdated information, which may no longer be useful. Discarding these packets prevents the transmission of stale information and contributes to reducing network congestion.

Following the definition, the Packet Error Rate ($\mathrm{PER}$) is computed as: 
\begin{equation}
    \mathrm{PER}_i = \frac{N_{\mathrm{dis},i}}{N_{\mathrm{gen},i}} \times 100,
\end{equation}

\noindent where $N_{\mathrm{dis},i}$ is the number of packets of slice $i$ discarded for exceeding the PDB, and $N_{\mathrm{gen},i}$ is the total number of packets generated in the slice $i$ during the observation period $\mathcal{T}$.
\subsubsection{eMBB Service Model}
In contrast to V2X services, eMBB traffic is modeled as a continuous data flow during $\mathcal{T}$. Users associated with these services continuously request data and remain active, thus persistently requiring radio resources at each TTI.

Following the packet arrival model from (\ref{lambda_service}), the eMBB service packet arrival rate is defined as in (\ref{lambda_service_embb}).

\begin{equation}
\label{lambda_service_embb}
\lambda^{\mathrm{embb}}_i
=
\frac{1}{|\mathcal{T}| \Delta t}
\sum_{t \in \mathcal{T} } A^{\mathrm{embb}}_i(t)
\end{equation}

The throughput experienced by eMBB users depends on the amount of radio resources allocated by the scheduler as well as on the channel conditions. Consequently, throughput is adopted as the main performance metric for eMBB services. Let $\mathrm{Th}^{\mathrm{embb}}_{i,u}(\mathcal{T})$ be the average throughput experienced by the user $u$ over the observation set $\mathcal{T}$, the performance of an eMBB service is characterized by 
:
\begin{equation}
\label{tput_service}
\mathrm{Th}^{\mathrm{embb}}_{i} = \frac{1}{|U^{\mathrm{embb}}_{i}|} \sum_{u \in U^{\mathrm{embb}}_i} {\mathrm{Th}}^{\mathrm{embb}}_{i,u}(\mathcal{T}),
\end{equation}

\noindent where $U^{\mathrm{embb}}_{i}$ is the set of users associated to the service and $|U_{i}^{\mathrm{embb}}|$ is its cardinality.

\subsection{Scheduler Description}

The radio resource assignment is performed in each TTI and, in an NS architecture, it can be logically divided into two hierarchical levels: inter-slice allocation and intra-slice resource scheduling. First, in inter-slice allocation, a subset $c_i$ of PRBs is assigned to each slice $n_{i}$ subject to the constraint in (\ref{constraint}), where $C_{\mathrm{tot}}$ represents the total number of PRBs available in the serving cell and $N$ the set of configured slices.

\begin{equation}
\label{constraint}
C_{\mathrm{occupied}} = \sum_{i \in N} c_i,
\quad \text{s.t.} \quad
C_{\mathrm{occupied}} \leq C_{\mathrm{tot}}
\end{equation}

After inter-slice resource allocation, intra-slice scheduling is performed independently for each slice according to the associated scheduling policy. In the scenario described, a Round Robin (RR) policy has been adopted to allocate the $c_i$ PRBs among the active users of the slice.

To improve resource efficiency, after the intra-slice scheduling, the remaining PRBs may be flexibly shared among users who still require them. These PRBs may arise from resources unassigned to any slice during the inter-slice allocation phase or from PRBs left unused within a slice due to over-provisioning. In both cases, the PRBs are scheduled independently of the service or slice to which the users are associated, following a RR policy.

\subsection{Problem Formulation For RAN Slicing}

This paper focuses on determining the optimal distribution of PRBs among network slices in a 5G RAN downlink scenario. The study considers four distinct services, each mapped to a dedicated slice, denoted as $n_i$. Specifically, slices $n_1$ and $n_2$ support V2X services, indexed by $\mathcal{I}_{\mathrm{v2x}} = \text{\{1, 2\}}$, while $n_3$ and $n_4$ are dedicated to eMBB services, indexed by $\mathcal{I}_{\mathrm{embb}} = {\{1, 2\}}$. Thereby, the inter-slice resource allocation is represented by the set $C = {\{c_1, c_2, c_3, c_4\}},$
where $c_i$ denotes the number of PRBs allocated to slice $n_i$.

Each V2X service is characterized by its packet size $s^{\mathrm{v2x}}_i$ and inter-packet generation period $T^{\mathrm{v2x}}_i$, allowing heterogeneous traffic patterns and intensities. In contrast, each eMBB service is associated with an expected throughput $\mathrm{Th}^{\mathrm{embb}}_{i,\mathrm{exp}}$.

The goal of the proposed inter-slice allocation problem is to determine the optimal PRB distribution $C$ such that:

\begin{itemize}
    \item V2X SLAs are strictly satisfied.
    \item Slice radio resource utilization is maximized.
    \item Remaining PRBs are allocated to eMBB slices according to their specific throughput needs.
\end{itemize}

These objectives are pursued under high-load conditions, where every slice receives at least 1 PRB and all available PRBs are allocated, such that $C_{\mathrm{occupied}} = C_{\mathrm{tot}}$.


Based on these assumptions, the inter-slice RAN slicing problem can be formulated as a multi-objective optimization:

\begin{equation}
\begin{aligned}
\max_{\mathbf{C}} \quad
& F(C) = f_{\eta}(C) + f_{\mathrm{embb}}(C) \\
\text{s.t.} \quad
& C_{\mathrm{tot}} = \sum_{i \in N} c_i , \\
& c_i \ge 1, \quad \forall i \in N, \\
& P_{\Gamma_i} \le D^{\mathrm{RAN}}_{\max,i},
\quad \forall i \in \mathcal{I}_{\mathrm{v2x}},
\end{aligned}
\label{problem_def}
\end{equation}

\noindent where $P_{\Gamma_i}$ denotes the 
percentile used to evaluate whether the SLA of V2X slice $i$ is satisfied. 

The objective function $F(C)$ jointly captures slice-level resource efficiency and eMBB throughput maximization. 
\begin{itemize}
    \item $f_\eta(C)$ accounts for the resource efficiency 
    in each slice $i$ during the observed time interval $\mathcal{T}$.
    \item $f_{\mathrm{embb}}(C)$ captures the average user throughput $\mathrm{Th}^{\mathrm{embb}}_{i}$ achieved in the eMBB slices.
\end{itemize}

This formulation defines a multi-objective optimization problem, where SLA compliance for V2X services acts as a hard constraint, while the slice resource efficiency and eMBB throughput maximization are optimized within the feasible set defined by the V2X SLA restriction.

\section{Latency Aware RAN Slicing Solution}
The RAN slicing problem aims to determine the optimal allocation of PRBs among the slices deployed in the network to satisfy the requirements of V2X services while minimizing the impact on overall performance. This problem can be addressed by a centralized controller at the cell level that determines the number $c_i$ of PRBs to be allocated to each slice. Such a decision must be made over time and adjusted to changing network conditions, including traffic demand and channel quality. This can be formulated as an adaptive decision-making process, which can be effectively addressed through RL, as it aims to find an optimal solution by dynamically interacting with the environment and learning from the consequences of decisions.

\subsection{DRL-Based Slicing Strategy}

The proposed approach involves a DRL agent that periodically determines the allocation of PRBs among the slices based on the performance of the services offered within the cell. To achieve this decision capability, during the training phase, the agent selects an action at every step, observes the resulting network behavior, and receives a reward that reflects the impact of the chosen action. Using these rewards, the DRL algorithm adjusts its decision-making process to learn which actions lead to the highest reward. 

To acquire this knowledge, the following elements are defined:

\subsubsection{State space} The state space defines the network characteristics observed by the agent before selecting an action. The state is represented by a vector $s_t$ of KPIs collected at the slice level and computed over an observation interval $\mathcal{T}$.

Specifically, the space vector comprises the following KPIs:

\begin{itemize}
    \item The number of PRBs allocated to each slice, $c_i$.
    \item The average resource utilization, defined as the average number of PRBs used by each slice, $c_{\mathrm{occupied},i}$.
    \item The average Signal-to-Interference-plus-Noise Ratio (SINR), $\mathrm{SINR}_i$, to capture the channel conditions experienced by users in each slice. 
    \item The service-level packet arrival rate $\lambda_i$, computed according to (\ref{lambda_service}) for V2X slices and (\ref{lambda_service_embb}) for eMBB slices, since each slice is dedicated to a single service.
    \item  The average number of active connections from a slice competing for radio resources, $q_i$, to capture the traffic demand and congestion level.
\end{itemize}

\subsubsection{Action space} The action space defines the possible allocation decisions that the agent can take. Since there are four slices deployed, this is defined as a vector $a = [a_1, a_2, a_3, a_4]$, where $a_i$ is the number of PRBs to be allocated to slice $i$. 

However, in the implemented algorithm, the action space is defined as a discrete set, and each action corresponds to an index that references a PRB allocation vector $a$ that satisfies the PRB budget constraint imposed in (\ref{problem_def}).

\subsubsection{Reward Computation} The reward obtained after executing an action informs how well the selected PRB allocation fulfills the objectives of the optimization problem. This reward is computed by the function proposed in (\ref{reward}), which follows a branched structure that allows the evaluation of two distinct scenarios.

\begin{equation}
R=
\begin{cases}

f_\eta(C)+f_{\mathrm{embb}}(C),
&\hspace{-6em} \text{if } P_i^{\mathrm{norm}}\le1,\ \forall i\in\mathcal{I}_{\mathrm{v2x}},\\[1ex]
\displaystyle
-\sum_{i\in\mathcal{I}_{\mathrm{v2x}}}
\left(
\gamma_i P_i^{\mathrm{norm}}
+\delta_i \mathrm{PER}_i^{\mathrm{norm}}
\right),
& \text{otherwise.}
\end{cases}
\label{reward}
\end{equation}

The first branch of the reward function is applied when the selected PRB allocation allows the fulfillment of the SLA in both V2X slices. This branch is designed to pursue the maximization of resource efficiency and eMBB user throughput from (\ref{problem_def}), once the percentile constraints are guaranteed.

The $f_\eta(C)$ function is defined in (\ref{prb_eff}) and computes the penalty for each slice $i$ as a negative reward weight $\alpha_i$ multiplied by the difference between $c_i$ and the average number of PRBs used in the slice, $c_{\mathrm{occupied},i}$, normalized by $c_i$. This formulation allows punishing over-provisioning while avoiding a slice from receiving more resources than actually needed. 

 Additionally, the factor $\frac{1}{\lambda_i}$ is introduced to adjust the penalty to the packet arrival frequency in each slice. Here, $\lambda_i$ represents the average packet arrival rate in the slice $i$, which corresponds to $\lambda^{\mathrm{v2x}}_i$ or to $\lambda^{\mathrm{embb}}_i$ depending on the service. For a given service with fixed packet sizes and transmission conditions, lower packet arrival rates present a larger set of possible PRB configurations that may satisfy the service requirements, including over-provisioning ones. The proposed weighting increases the penalty for resource-inefficient allocations in these cases, encouraging the selection of more efficient configurations among those that meet the service requirements.  
 \begin{equation}
    f_\eta(C) = \sum_{i \in \mathcal{N}} f_{\eta,i}(C) = \sum_{i \in \mathcal{N}} \alpha_i \frac{1}{\lambda_i}\max\!\left(0, \frac{c_i - c_{\mathrm{occupied},i}}{c_i}\right)
    \label{prb_eff}
\end{equation}

 The second reward term, $f_{\mathrm{embb}}(C)$, is defined in (\ref{reward_tput}) and accounts for the throughput achieved by the eMBB users from each slice. Specifically, it incentivizes PRB allocations that increase the average user throughput from (\ref{tput_service}), which is normalized by the expected user throughput. This enables focusing on how close the service of each slice operates to its performance expectations rather than its raw throughput.
 \begin{equation}
    f_{\mathrm{embb}}(C) = \sum_{i \in \mathcal{I}_{\mathrm{embb}}}\rho_i\min\!\left(1, \frac{Th^{\mathrm{embb}}_{i}}{{Th}^{\mathrm{embb}}_{i,\mathrm{exp}}} \right)
\label{reward_tput}
\end{equation}
 
The parameter $\rho_i$ in $f_{\mathrm{embb}}(C)$ adjusts the component of each slice according to the service's expected throughput. 
In scenarios with heterogeneous performance expectations, the agent might tend to optimize the slice that requires fewer resources to achieve high performance; by scaling the contribution of each eMBB slice according to (\ref{ec:rho}), a more balanced optimization is promoted, as this prevents the agent from systematically prioritizing services that are easier to satisfy.

\begin{equation}
\rho_i =
\frac{Th^{\mathrm{embb}}_{i,\exp}}
     {\displaystyle\sum_{\substack{j \in \mathcal{I}_{\mathrm{embb}} \\ j \neq i}}
      Th^{\mathrm{embb}}_{j,\exp}},
\qquad
i \in \mathcal{I}_{\mathrm{embb}}
\label{ec:rho}
\end{equation}

In line with the delay constraint introduced in (\ref{problem_def}), the second branch computes the reward when the proposed PRB allocation does not allow all slices to meet the V2X service SLA, i.e, when the condition $P_{\varGamma, i} \leq D^{\mathrm{RAN}}_{\mathrm{max}}$ is not satisfied. In such circumstances, a penalty is received through the reward function. 
To capture the severity of the SLA violation, the penalty is composed of the following two terms:
\begin{itemize}
    \item The aggregated $P^{\mathrm{norm}}_i$ components defined in (\ref{compo}), where $P_{\varGamma, i}$ is normalized by its maximum tolerable delay, $D^{\mathrm{RAN}}_{\max,1}$. This enables a relative assessment of the degradation experienced by the V2X service in each slice and penalizes the action proportionally to the magnitude of the SLA violation.
    \item The aggregated $\mathrm{PER}^{\mathrm{norm}}_i$ components defined in (\ref{compo}), where $\mathrm{PER}_i$ is normalized to a scale of 0 to 1. This indicator represents the proportion of packets discarded due to exceeding the PDB. 
\end{itemize}

Finally, the parameters $\gamma_i$ and $\delta_i$ are reward weights and control the contribution of each term in the reward function.

This definition allows penalizing PRB distributions in proportion to both the delay degradation and the packet loss, imposing stronger penalties as the severity of the SLA violation increases.

\begin{equation}
P^{\mathrm{norm}}_i = \frac{P_{\varGamma,i}}{D^{\mathrm{RAN}}_{\max,i}},
\quad
\mathrm{PER}^{\mathrm{norm}}_i = \frac{\mathrm{PER}_i}{100}
\label{compo}
\end{equation}
 
\subsubsection{PPO Agent}

Among the various DRL algorithms, the solution is based on Proximal Policy Optimization (PPO), an actor-critic implementation from \cite{schulman}, in which two neural networks are used. The first one represents the policy $\pi_{\theta}(a|s_t)$ and corresponds to the actor, which outputs a probability distribution over the possible actions $a$ given the current state $s_t$ at each time step $t$. The second one is the critic, and estimates the value function $V^{\pi}(s_t)$ associated with the current $\pi_{\theta}$ policy. In this work, both the actor and critic are modeled as Multi-Layer Perceptrons (MLP) with two hidden layers and tanh activation function.

According to its definition \cite{schulman}, to prevent abrupt policy updates and add stability to the algorithm, a clipping mechanism is introduced in the loss function $L_{t}(\theta)$. This implementation also reduces the variance of the policy updates and improves convergence by relying on Generalized Advantage Estimation (GAE) to estimate the advantage $\hat{A}_t$ \cite{schulman}.


 


\subsubsection{Environment and Training Methodology}

As highlighted in \cite{cita_an_efficent}, solving such optimization problems in real deployments is highly complex due to dynamic traffic demand, heterogeneous service requirements, and time-varying channel conditions. Furthermore, online training could negatively impact the user experience, especially during the early exploration stages. To overcome these drawbacks, the agent operates in an environment built from pre-collected network observations stored in a dataset $\mathcal{D}$. This dataset comprises observations under different PRB allocations ($C$) and load conditions, which will be identified by a load level $L$ according to the packet arrival rates $\lambda_i$. In this way, state transitions are modeled using the dataset, allowing the agent to learn the consequences of applying each action without requiring direct interaction with a real deployment.


\begin{figure*}
    \centering
    \includegraphics[width=1\linewidth]{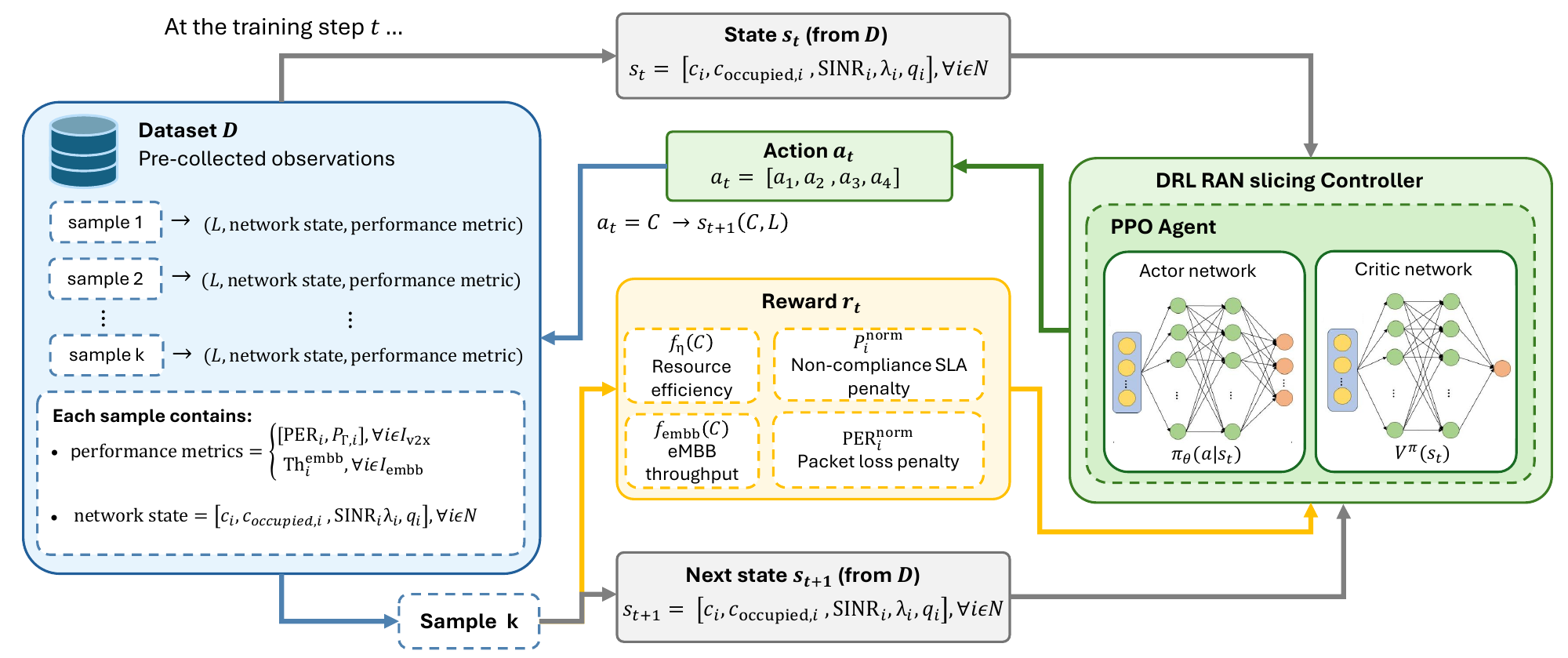}
    \caption{Model training process within an episode.}
    \label{fig:rl_train}
\end{figure*}

As illustrated in Algorithm \ref{alg:ppo_prb}, the training process is organized into episodes, and the initial state of each episode is determined by two factors: 
\begin{itemize}
    \item \textbf{Episode Load Level $L$:} To promote adaptability to varying traffic conditions, the load level $L$ is randomly selected at the beginning of each episode. Once selected, it remains fixed throughout the episode to ensure stationary load conditions.
    
    \item \textbf{The initial PRB allocation:} This is selected following a worst-case strategy where the initial PRB allocation is chosen from the set of configurations that yield the highest aggregated $P_{\varGamma, i}$ across all V2X slices. This encourages the agent to focus on critical delay conditions.
\end{itemize}

As a result, the initial network state is obtained by sampling an entry from the dataset $D$, featuring the selected load level $L$ and the resource allocation $C$ chosen based on the worst-case criterion.

Once the initial state has been defined, training proceeds as shown in Fig. \ref{fig:rl_train}. At each training step $t$, a network state $s_t$ is observed. This state is retrieved from the dataset $D$. Based on this state, the actor network samples an action $a_t$, which corresponds to a new PRB allocation $C$. Based on this PRB allocation and the load level $L$ in the observed state $s_t$, a new sample from the dataset $D$ is selected to constitute the next state $s_{t+1}$. From this entry, performance metrics including $P^{\mathrm{norm}}_i$, $\mathrm{PER}_i$ and $\mathrm{Th}_i^{\mathrm{embb}}$ will be extracted to compute the resulting reward. Therefore, the transition from $s_t$ to $s_{t+1}$ reflects the impact of selecting a PRB allocation preserving the underlying traffic conditions.

\begin{algorithm}
\caption{PRB Allocation Based on PPO}
\label{alg:ppo_prb}
\begin{algorithmic}[1]

\STATE \textbf{Inputs:}
\STATE $N$: Number of slices
\STATE $C_{\mathrm{tot}}$: Total number of PRBs available in the cell
\STATE $\mathcal{D}$: Dataset with precomputed KPIs
\STATE $SLA^{\mathrm{v2x}}_i = (D^{\mathrm{RAN}}_{\mathrm{max},i}, \Gamma_i), \forall i \in \mathcal{I}_{\mathrm{v2x}}$
\STATE $Th^{eMBB}_{i,max}, \forall i \in \mathcal{I}_{\mathrm{eMBB}}$
\STATE $strategy$: "worst case"

\STATE \textbf{Initialize} actor network $\pi_\theta$ and critic network $V_\phi$
\STATE \textbf{Initialize} experience buffer $\mathcal{B}$

\WHILE{training not finished}

    \STATE Select load level $\rightarrow$ $L$ 
    \STATE Select initial PRB allocation($strategy$, $L$) 
    
    \WHILE{episode not terminated}

        \STATE Observe state $s_t$ retrieved from $\mathcal{D}$
        \STATE Sample action $a_t \sim \pi_\theta(a_t|s_t)$
        
        \STATE Apply action $a_t$ 
        \STATE Retrieve next state $s_{t+1}$ from dataset $\mathcal{D}$
        \STATE Compute reward $r_t$ 
        
        \STATE Store transition $(s_t, a_t, r_t, s_{t+1})$ in buffer $\mathcal{B}$
        \IF{Update step is reached}
            \STATE Compute advantage $\hat{A}_t$ using GAE
            \STATE Perform PPO update
        \ENDIF
        \STATE $t \leftarrow t+1$
        
    \ENDWHILE

\ENDWHILE

\end{algorithmic}
\end{algorithm}

\section{DRL Slicing Policy Evaluation}

\subsection{Simulation Setup}
As described in the previous section, agent training relies on a dataset of network observations that encompasses different traffic load levels and resource partitioning configurations. In this case, these observations are generated through a 5G system-level simulator with RAN Slicing capability.

The considered scenario corresponds to a cell from a tri-sectorized site as the one presented in Fig. \ref{fig_ML_cycle} since the optimization is performed at the cell level. The main simulation parameters are summarized in Table \ref{tab:tablecaracs}, where the bandwidth is specified as the total number of available PRBs, $C_{\mathrm{tot}} = 50$.

Two V2X services are defined, each representing a different vehicular traffic pattern adapted from \cite{trafico3GPP} to meet simulator constraints. The first V2X service, hereafter denoted as V2X-1 with $i=1$, models a medium-intensity periodic traffic with an inter-packet arrival time of 10ms and a packet size of 6400 bits. The second V2X service, V2X-2 with $i=2$, corresponds to a lower traffic intensity model, where $T^{\mathrm{v2x}}_{2}$ is set to 50ms and $s^{\mathrm{v2x}}_{2}$ to 2400 bits. 
It is important to note that, even though packets arrive periodically for each user, users are not uniformly distributed over time; therefore, the instantaneous demand fluctuates across time steps.

In addition, two eMBB services are defined, each characterized by different expected rates. The first eMBB service, denoted as eMBB-1 with $i=1$, is configured with a $\mathrm{Th}_{1,\mathrm{exp}}^\mathrm{embb}$ of 2.24Mbps, corresponding to the user throughput achievable with 1 PRB under favorable channel conditions.
The second service, eMBB-2 with $i=2$ is designed to be more demanding, with a $\mathrm{Th}_{2,\mathrm{exp}}^\mathrm{embb}$ of 11.2Mbps, which corresponds to the throughput achievable with 5 PRBs under the same conditions.

It should be noted that both $\mathrm{Th}_{1,\mathrm{exp}}^\mathrm{embb}$ and $\mathrm{Th}_{2,\mathrm{exp}}^\mathrm{embb}$ represent a reference value rather than a upper bound. 
\begin{table}[t]
\caption{Simulation Parameters}
\label{tab:tablecaracs}
\centering
\begin{tabular}{|l|p{0.5\columnwidth}|}
\hline
\textbf{Parameter} & \textbf{Value} \\
\hline
\multicolumn{2}{|c|}{\textbf{Network Parameters}} \\
\hline
PRBs per cell ($C_{\mathrm{tot}}$) & 50 \\
Power transmission & 36 dBm \\
Gain transmission & 20 dBi \\
Carrier frequency & 3.5 GHz \\
ScS & 15KHz \\
Packet Delay Budget & 30 ms \\
TTI duration & 1 ms \\
Base station height & 25 meters \\
Pathloss model & Pathloss and LOS probability modeled as in \cite{pathloss} \\
\hline
\multicolumn{2}{|c|}{\textbf{V2X-1 Parameters}} \\
\hline
Inter-packet arrival period & $T^{\mathrm{v2x}}_{1} = 10$ ms \\
Packet size & $s^{\mathrm{v2x}}_{1} = 6400$ bits \\
UE speed & 50 km/h \\
\hline
\multicolumn{2}{|c|}{\textbf{V2X-2 Parameters}} \\
\hline
Inter-packet arrival period & $T^{\mathrm{v2x}}_{2} = 50$ ms \\
Packet size & $s^{\mathrm{v2x}}_{2} = 2400$ bits \\
UE speed & 50 km/h \\
\hline
\multicolumn{2}{|c|}{\textbf{eMBB-1 Parameters}} \\
\hline
User throughput expectations & $Th^{\mathrm{embb}}_{1, \mathrm{exp}} = 2.24$ Mbps \\
UE speed & 4 km/h \\
\hline
\multicolumn{2}{|c|}{\textbf{eMBB-2 Parameters}} \\
\hline
User throughput expectations& $Th^{\mathrm{embb}}_{2, \mathrm{exp}} = 11.2$ Mbps \\
UE speed & 4 km/h \\
\hline
\end{tabular}
\end{table}

Each service is associated with a dedicated network slice $n_i$. Therefore, the slice assignment is:
\begin{equation}  
\label{eq:slice_conf}
\text{\{$n_1, n_2, n_3, n_4$}\} = \{\text{V2X-1}, \text{V2X-2}, \text{eMBB-1}, \text{eMBB-2}\}\
\end{equation}


\begin{table*}[h]
\caption{Packet generation rates and number of users for each load level.}
\centering
\begin{tabular}{|c|c|c|c|c|c|c|c|c|}
\hline
\textbf{$L$} &
$\bar{\lambda}^{\mathrm{v2x}}_1 \pm \sigma$ &
$\bar{\lambda}^{\mathrm{v2x}}_2 \pm \sigma$ &
$\bar{\lambda}^{\mathrm{embb}}_1 \pm \sigma$ &
$\bar{\lambda}^{\mathrm{embb}}_2 \pm \sigma$ &
$U^{\mathrm{v2x}}_1$ &
$U^{\mathrm{v2x}}_2$ &
$U^{\mathrm{embb}}_1$ &
$U^{\mathrm{embb}}_2$ \\
\hline
1 & 1.82 $\pm$ 0.09 & 1.55 $\pm$ 0.06 &
13.43 $\pm$ 0.40 & 13.29 $\pm$ 0.44 &
15 & 50 & 15 & 15 \\
\hline
2 & 3.03 $\pm$ 0.09 & 2.18 $\pm$ 0.06 &
8.87 $\pm$ 0.37 & 9.41 $\pm$ 0.35 &
30 & 100 & 10 & 10 \\
\hline
3 & 6.02 $\pm$ 0.13 & 4.00 $\pm$ 0.01 &
9.10 $\pm$ 0.31 & 8.07 $\pm$ 0.53 &
60 & 200 & 10 & 10 \\
\hline
\end{tabular}
\label{packets/ms}
\end{table*}

To validate the proposed solution, three main load levels have been considered. 
As summarized in Table \ref{packets/ms}, each of them can be characterized by the mean and standard deviation of the packet generation rates across all simulations under a load level. As a result, three operating scenarios corresponding to low, medium, and high demand conditions are defined. They are expressed in packets/ms, since 1ms is the minimum temporal resolution of the simulation tool. The number of users configured for each service, denoted as $U_{i}^{\mathrm{v2x}}$ for $i \in \mathcal{I}_{\mathrm{v2x}}$ and $U_{i}^{\mathrm{embb}}$ for $i \in \mathcal{I}_{\mathrm{embb}}$, is also provided. These values were selected to set a broad range of V2X resource requirements, ranging from scenarios in which the V2X slices require only a small fraction of the available resources to highly demanding scenarios requiring a significantly larger allocation.




Overall, the resulting dataset is constructed from simulations executed under these load levels and a representative set of PRB allocation configurations to capture different operating conditions. Since each simulation is very time-consuming, the action space has been discretized into 47 PRB configurations. The selected configurations span meaningful operating points, with PRB allocations ranging from 1 to 25 for V2X-1, 1 to 13 for V2X-2, 3 to 24 for eMBB-1, and 1 to 44 for eMBB-2. These configurations include allocations that lead to SLA violations, allocations that satisfy the V2X requirements, and overprovisioning configurations for one or both V2X slices while varying the resources assigned to the eMBB slices. Each PRB allocation $C$ has been simulated under the three selected load levels, and multiple simulation samples have been generated for every combination of $C$ and $L$ to capture their impact under different traffic conditions. In total, the resulting training dataset comprises 564 samples.

From these simulations, each with an observation interval $\mathcal{T}$ of 30000 TTIs (30 seconds), the KPIs are computed at the slice level. This observation interval $\mathcal{T}$ has been selected along with the traffic patterns from Table \ref{tab:tablecaracs} and the load levels as a trade-off between simulation time and statistical representativeness, since this configuration provides a sufficiently large number of packet transmissions to obtain stable KPIs.

Finally, the experienced delay of a V2X packet, $D_{i}^{\mathrm{v2x}}$, is defined as the elapsed time between its arrival at the scheduler and its termination. For successfully transmitted packets, this termination occurs when the last bit of the packet is transmitted, whereas for packets that do not complete their transmission before the PDB expires, this occurs when the packet is discarded. Since the simulator operates with a time resolution of 1ms, a packet that exceeds the PDB will be identified at the following TTI boundary and recorded with an experienced delay of PDB+1 ms. 

Both successfully transmitted and discarded packets are included in the packet-delay distribution to characterize the delay experienced by the overall set of generated packets and to avoid overly optimistic $P_{\varGamma, i}$ delay percentiles. 


\subsection{Training performance Analysis}

The proposed optimization strategy is evaluated under the two SLAs presented in Table \ref{tab:sla_reqs}. The maximum packet delay, $D^{\mathrm{RAN}}_{\mathrm{max}}$, and reliability, $\varGamma$, are derived from the requirements outlined in \cite{requirements3GPP} and the portion reserved for the core network according to the 5G QoS indicators (5QI) table from \cite{5QI_table}.
\begin{table}[t]
\caption{V2X SLA definition\label{tab:sla_reqs}}
\centering
\begin{tabular}{|c|c|c|}
\hline
\textbf{SLA} & \textbf{Maximum packet delay} & \textbf{Reliability} \\
\hline
1 & 15ms & 90\% \\
\hline
2 & 6ms & 99.99\% \\
\hline

\end{tabular}
\end{table}

Each SLA is evaluated in an independent experiment, in which both V2X slices have to satisfy the same SLA. Accordingly, two models have been trained, one for SLA 1 and another for SLA 2, each aiming to ensure that both slices achieve the required $P_{\varGamma, i}$ for SLA compliance.

\begin{figure*}[h]
\centering
\subfloat[]{\includegraphics[width=3.5in]{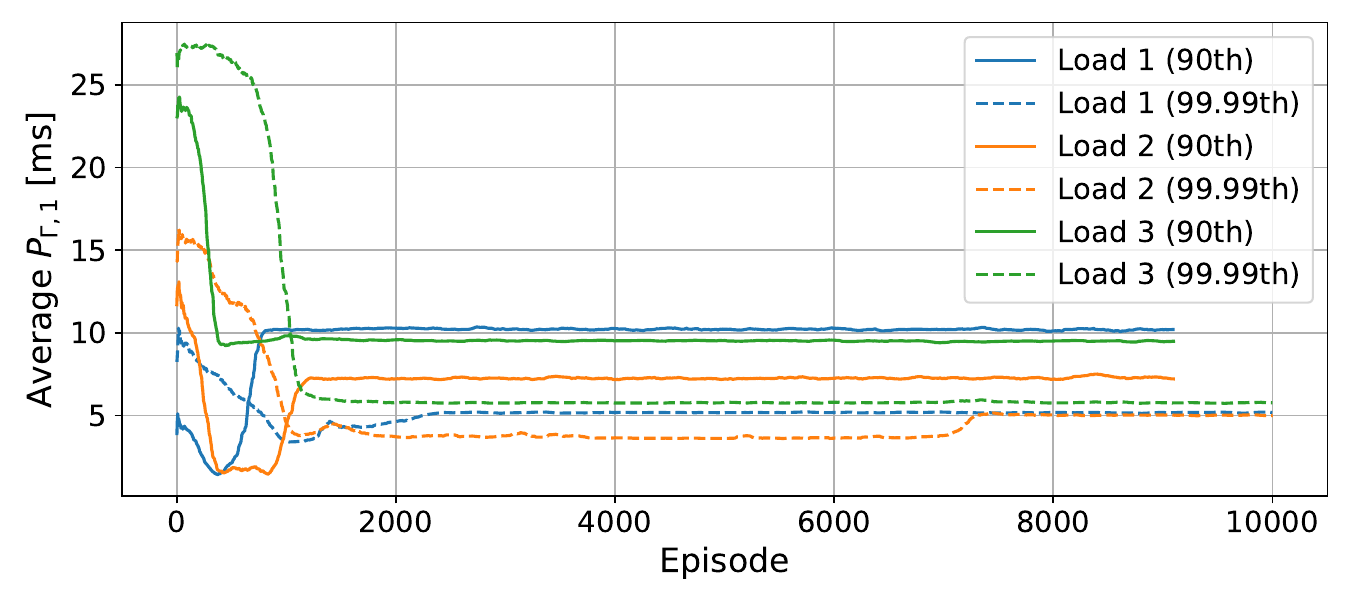}%
\label{v2x1_per}}
\hfil
\subfloat[]{\includegraphics[width=3.5in]{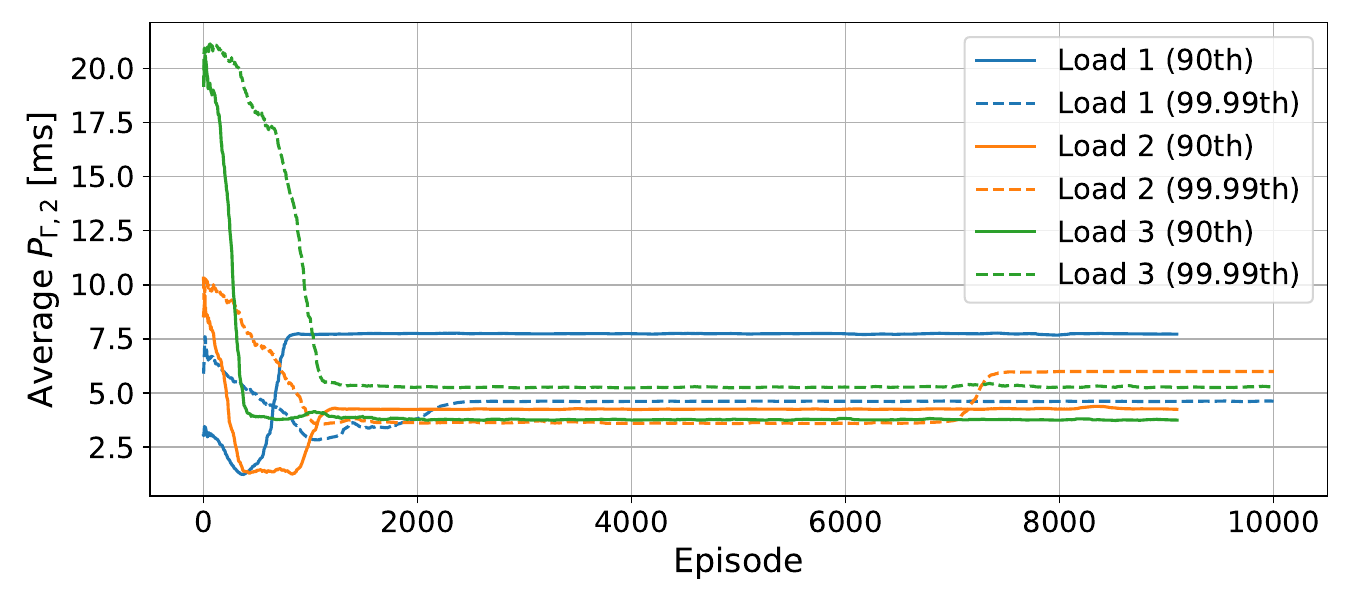}%
\label{v2x2_per}}
\hfil
\subfloat[]{\includegraphics[width=3.5in]{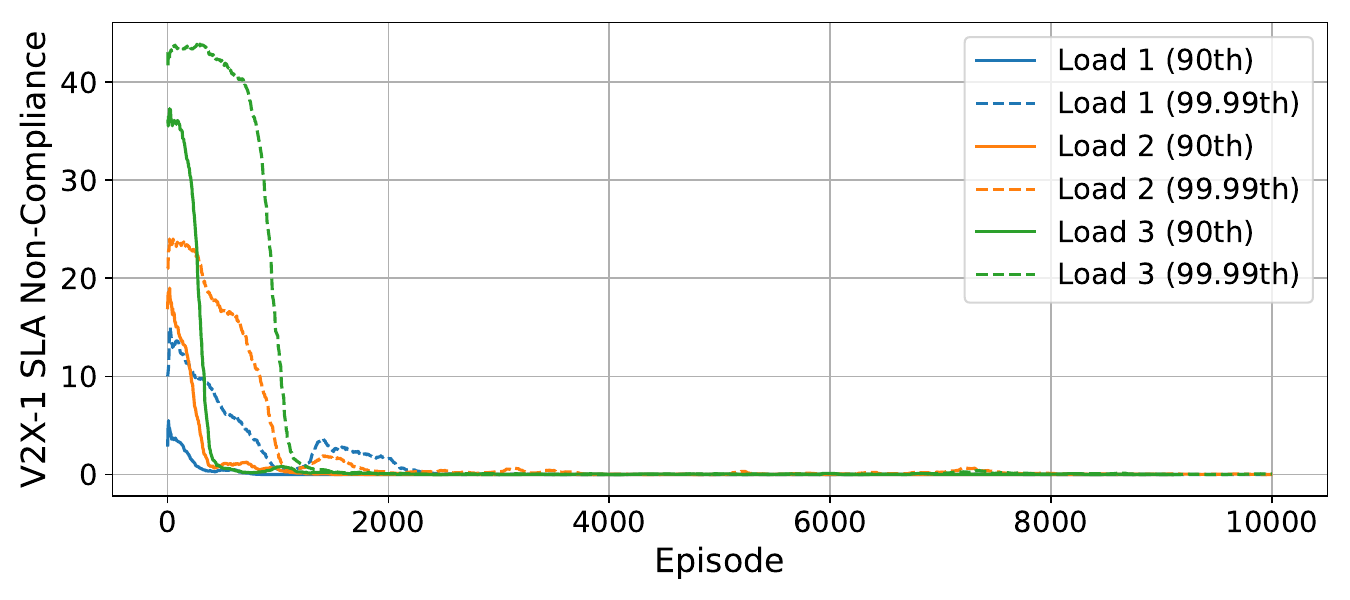}%
\label{v2x1_sla}}
\hfil
\subfloat[]{\includegraphics[width=3.5in]{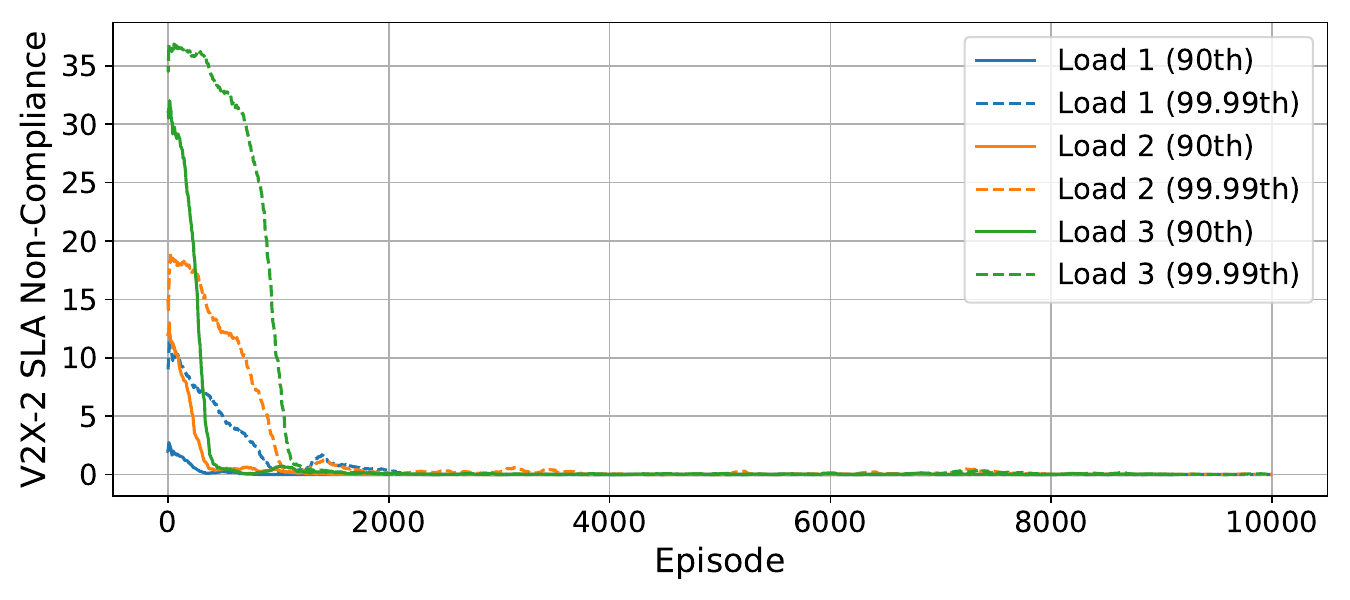}%
\label{v2x2_sla}}

\caption{Episode-averaged 90-th and 99.99-th percentile of radio delay and SLA compliance in slices dedicated to V2X services during training. (a) Episode-averaged $\varGamma$-th percentile of radio delay in V2X-1 slice. (b) Episode-averaged $\varGamma$-th percentile of radio delay in V2X-2 slice. (c) SLA Non-Compliance Steps per Episode in V2X-1 slice.(d) SLA Non-Compliance Steps per Episode in V2X-2 slice.}
\label{fig_sla}
\end{figure*}

For each experiment, the agent is trained under different load levels applying the corresponding parameters from Table \ref{tab:hiperparams}, to obtain a slicing policy. The table includes the reward weights $\gamma_{i}$, $\delta_{i}$ and $\alpha_{i}$, together with the PPO hyperparameters. These hyperparameters, including the learning rate, which sets the rate at which the model's parameters are updated during training, have been selected through an optimization process based on Optuna \cite{optuna}. The resulting policies are analyzed in terms of SLA compliance in Fig \ref{fig_sla}.

\begin{table}[t]
\caption{Model parameters\label{tab:hiperparams}}
\centering
\begin{tabular}{|c|c|c|}
\hline
\textbf{Parameter} & \textbf{SLA 1} & \textbf{SLA 2} \\
\hline
Learning rate & 0.00019 & 0.00028 \\
\hline
Steps to run before next update & 512 & 256 \\
\hline
Optimization epochs & 10 & 10 \\
\hline
Minibatch size & 32 & 64 \\
\hline
Discount factor & 0.9003 & 0.9076 \\
\hline
Factor for trade-off of bias vs variance for GAE & 0.9235 & 0.9707 \\
\hline
Clipping parameter & 0.3 & 0.1 \\
\hline
Entropy coefficient for the loss calculation & 0.005 & 0.01 \\
\hline
Neurons per layer & 32 & 64 \\
\hline
$\gamma_i,\delta_i, \ \forall i\in\mathcal{I}_{\mathrm{v2x}}$ & -40 & -40 \\
\hline

$\alpha_{i} ,\forall i\in\mathcal{N}$ & -40 & -40 \\

\hline
\end{tabular}
\end{table}





Figures \ref{v2x1_per} and \ref{v2x2_per} capture the average 90th and 99.99th percentiles of packet delay achieved per training episode. 

In the early stages, the average $P_{\varGamma, i}$ for the services in each slice exhibits variability. For SLA 1, high traffic loads show a large deviation from the required value during the initial episodes. In contrast, at the lowest level, a wide set of PRB distributions can satisfy the SLA, in some cases excessively, leading to lower average percentiles during early episodes. Around episode 400, the model begins to discard actions that violate the SLA, resulting in a noticeable drop in the average percentile. At this stage, the policy focuses on SLA-compliant actions, many associated with over-provisioning, which contributes to the low percentiles observed. After some episodes, the average percentile moves away from over-provisioning and stabilizes below the 15ms threshold, demonstrating that the learned policy successfully meets the SLA. 

A similar trend is observed for the 99.99th percentile. In the early stages, 
 the average percentile decreases significantly, 
 since fewer PRB allocations allow it to be met. After 7000 episodes, the 99.99th percentile of all load levels converges to values around or below 6ms, confirming that the learned policy allows achieving the performance targets.

Additionally, Figures \ref{v2x1_sla} and \ref{v2x2_sla} display the number of SLA violations per episode. Both figures show that the learned policy minimizes SLA violations for both services across all traffic loads. Initially, violation rates are high, but as the training progresses, they decrease and become isolated events, indicating that the models effectively learn to meet the SLA.

\begin{figure*}[h]
\centering
\subfloat[]{\includegraphics[width=3.5in]{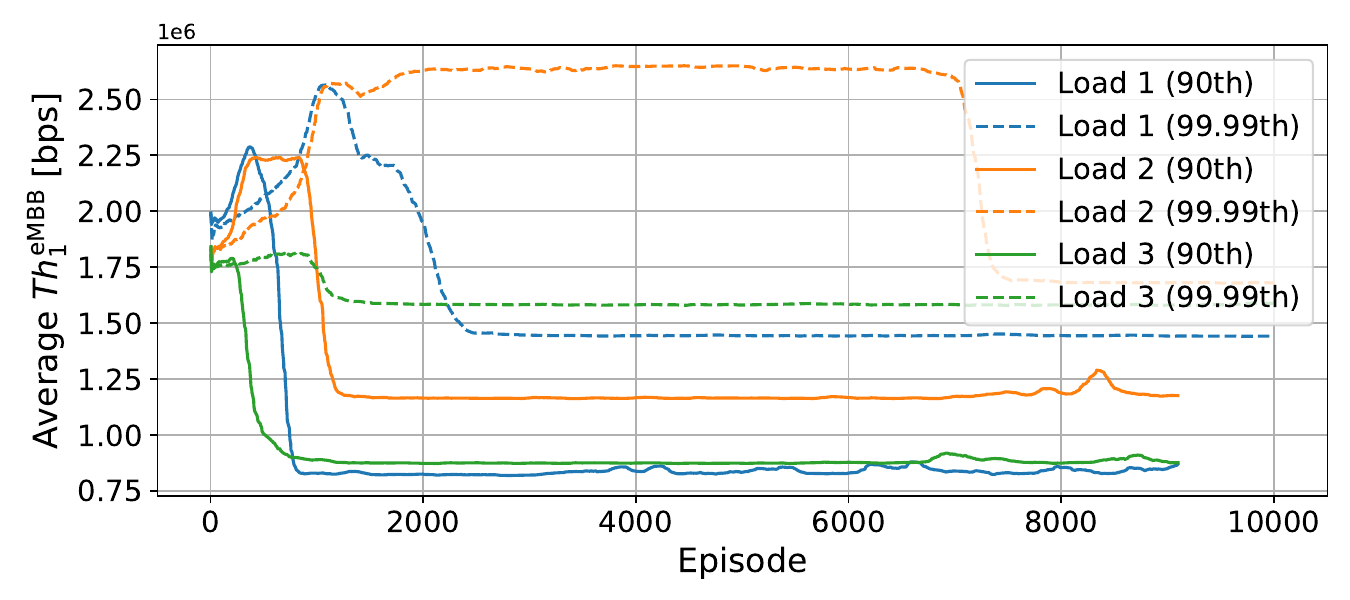}%
\label{embb1_tput}}
\hfil
\subfloat[]{\includegraphics[width=3.5in]{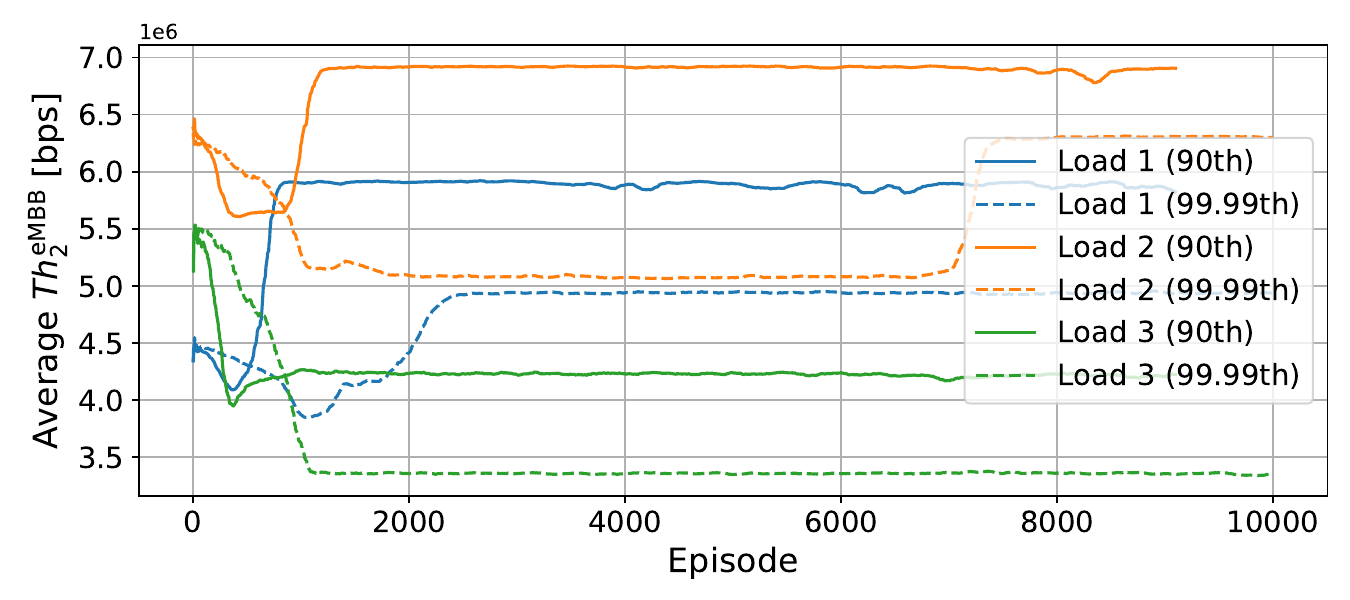}%
\label{embb2_tput}}
\hfil

\caption{Episode-averaged user throughput achieved during training. (a) Episode-averaged user throughput in the eMBB-1 slice. (b) Episode-averaged user throughput in the eMBB-2 slice.}
\label{fig_distribuciones}
\end{figure*}

In addition to ensuring compliance with the SLA for V2X services, the optimization strategy also aims to maximize eMBB service throughput. In slice eMBB-1, the episode-averaged user throughput increases during the first episodes, while slice eMBB-2 exhibits the opposite behavior. As observed in Figures \ref{embb1_tput} and \ref{embb2_tput}, after several episodes, both slices converge to a more balanced operating point. This behavior is closely related to the evolution of V2X percentiles. During episodes of V2X slice over-provisioning, some of the resources reserved for those segments are unused and can be opportunistically exploited by eMBB slices. This particularly benefits the eMBB-1 service, whose expected user throughput of 2.24 Mbps can be achieved in the simulator with a single PRB under favorable transmission conditions. However, variation in channel conditions may reduce the transmission efficiency. In such cases, allocating more than 1 PRB to eMBB-1 users can compensate for the reduced transmission efficiency, but it may also result in a user throughput above the expected value. When unused V2X resources are available, they can be allocated to eMBB-1 more frequently, increasing the probability of exceeding its expected throughput, which explains the throughput values observed for eMBB-1. 

Although the current policy satisfies the V2X constraints and yields high throughput in eMBB-1, it creates an imbalance between the eMBB services. As training progresses, the agent learns a fairer allocation policy, reducing the episode-averaged user throughput of eMBB-1 while increasing that of eMBB-2.


Interestingly, with a 99.99\% SLA, eMBB-1 achieves higher throughput values than with a 90\% SLA, whereas eMBB-2 records lower values. Although stricter SLAs require reserving more PRBs for V2X and these slices are dimensioned to guarantee good performance even during peak traffic periods, the periodicity of the V2X traffic means they are not fully utilized at every time instant. Consequently, a larger set of reusable PRBs is created compared to the 90\% SLA case, where fewer PRBs are dedicated to V2X slices, and more can be reserved to eMBB-2. In these scenarios, eMBB-1 benefits from this dynamic reuse, as it requires fewer resources to achieve relatively high performance. 

\begin{figure*}[h]
\centering
\subfloat[]{\includegraphics[width=3.5in]{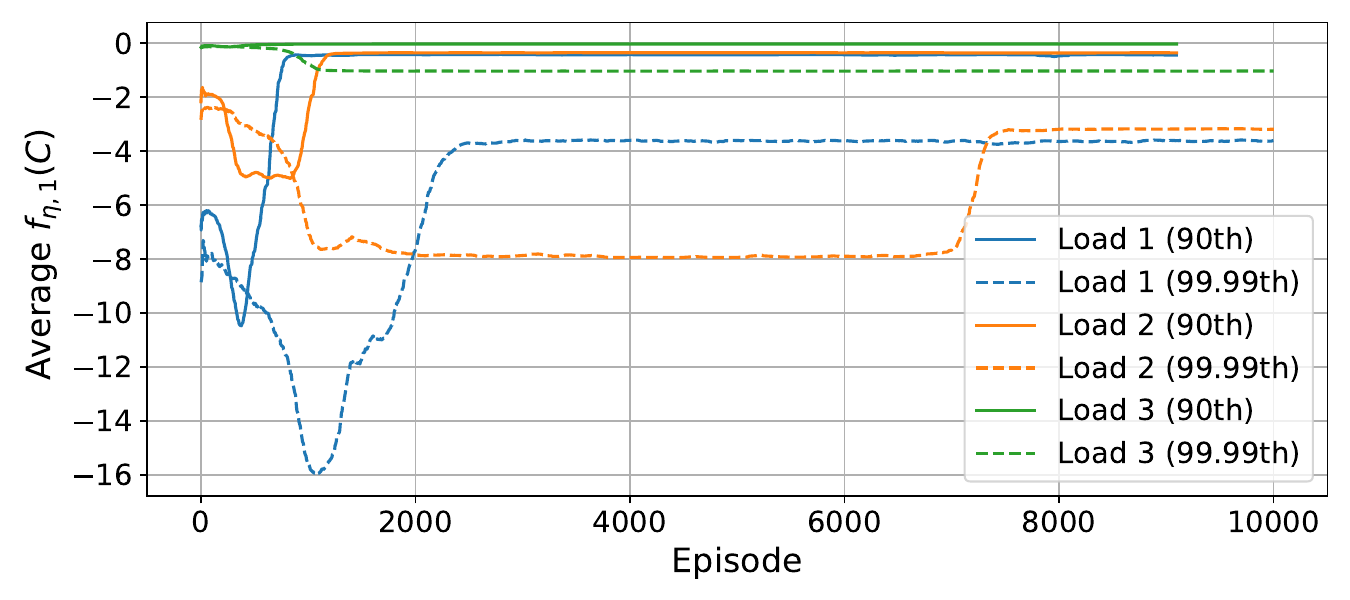}%
\label{prbs_v2x1}}
\hfil
\subfloat[]{\includegraphics[width=3.5in]{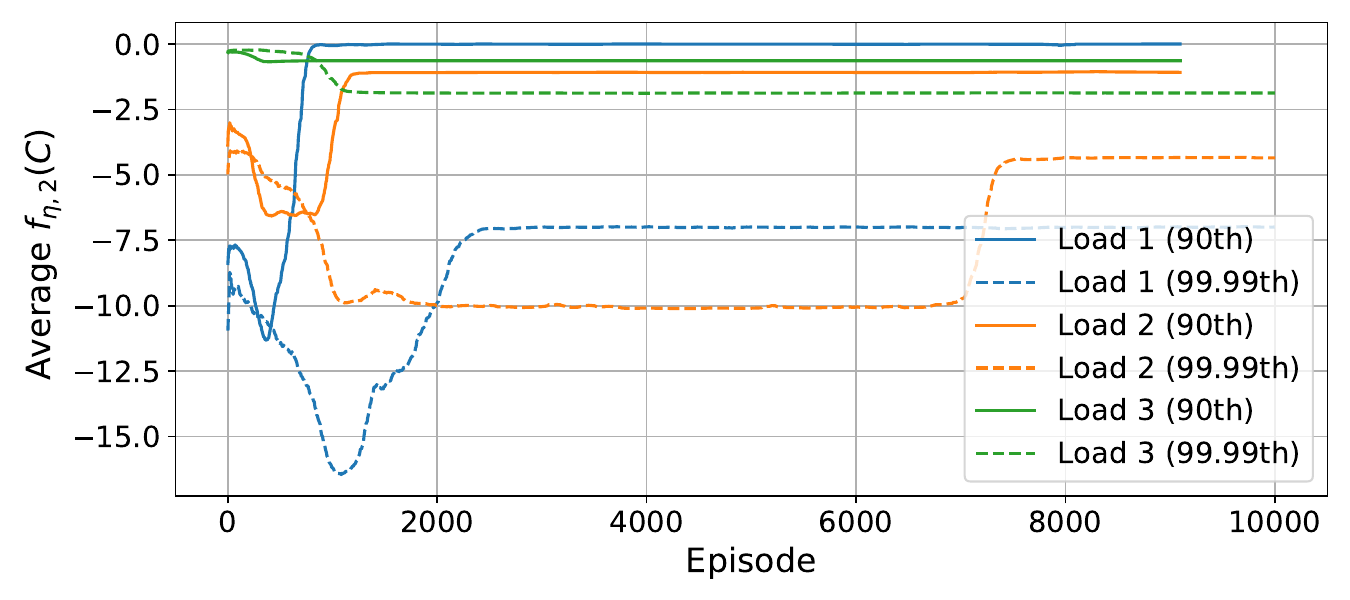}%
\label{prbs_v2x2}}
\hfil
\subfloat[]{\includegraphics[width=3.5in]{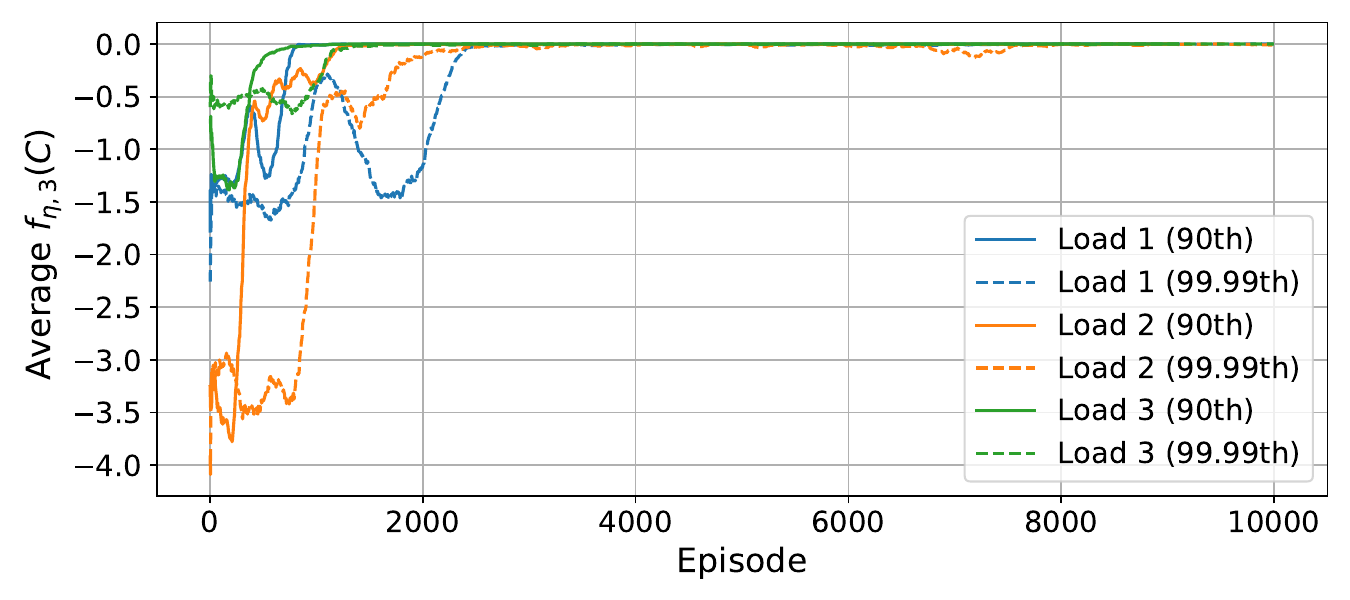}%
\label{prbs_embb1}}
\hfil
\subfloat[]{\includegraphics[width=3.5in]{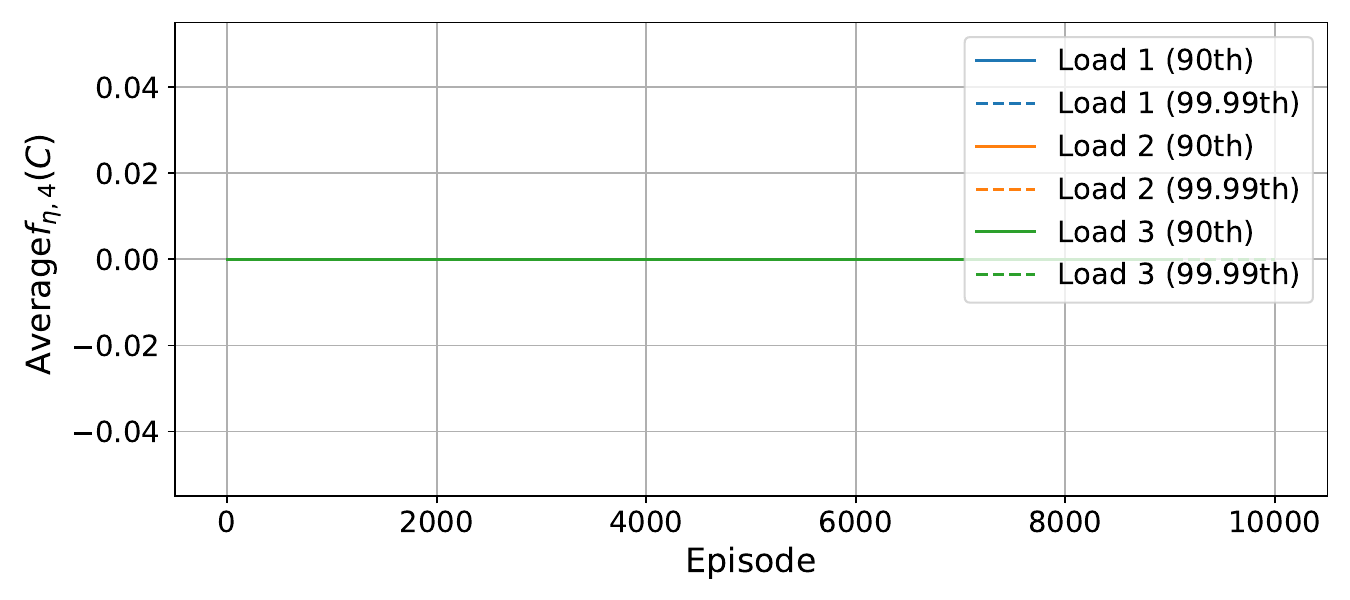}%
\label{prbs_embb2}}
\hfil
\caption{Average $f_{\eta,i}(C)$ component from reward function during training. (a) Average $f_{\eta,1}(C)$ per episode in the slice dedicated to the V2X-1 service. (b) Average $f_{\eta,2}(C)$ per episode in the slice dedicated to the V2X-2 service. (c) Average $f_{\eta,3}(C)$ per episode in the slice dedicated to the eMBB-1 service.(d) Average $f_{\eta,4}(C)$ per episode in the slice dedicated to the eMBB-2 service.}
\label{fig_prbs}
\end{figure*}

Additionally, an abrupt change in the average throughput is observed at Load level 2 under SLA 2. In Fig. \ref{embb1_tput} the throughput of eMBB-1 decreases from approximately 2.65 Mbps to 1.6Mbps, while the eMBB-2 throughput increases from 5.08 to 6.3 Mbps, according to Fig. \ref{embb2_tput}. This behavior is a consequence of a PRB allocation policy shift that can be better understood by examining Fig. \ref{fig_prbs}. 

Finally, resource efficiency is captured in Fig. \ref{fig_prbs} through the episode-averaged $f_{\eta}(C)$ component of each slice, denoted as $f_{\eta,i}$ where $i$ refers to a specific slice. Fig. \ref{prbs_v2x1} and Fig. \ref{prbs_v2x2} illustrate the relative gap between the PRBs allocated to V2X slices and the average PRBs actually utilized. 

During the first episodes, a significant reduction is observed, especially in the first two load levels. In these scenarios, a large number of configurations lead to over-provisioning, which contributes to reaching high negative values during the first episodes, where the learned policy focuses on meeting the SLA without accounting for resource efficiency. Meanwhile, for the third load level, the curve achieves less negative values since the higher demand reduces the action space that leads to V2X slice over-provisioning.

However, as the training progresses, the gap between allocated and utilized PRBs is reduced, increasing the values achieved by the component $f_{\eta,i}$ during an episode. Eventually, this component stabilizes, demonstrating that the agent has converged to a policy that not only ensures SLA compliance but also seeks to use resources efficiently. This is particularly evident for Load level 2 under SLA 2, where a policy update around episode 7000 leads to a smaller gap between allocated resources and utilized resources, as fewer PRBs are now dedicated to V2X services. This adjustment explains the abrupt change observed in the eMBB throughput, since now fewer PRBs remain available to be utilized by eMBB services. As a consequence, eMBB-1 has fewer opportunities to exploit unused V2X PRBs, while a larger share of the resources can be explicitly allocated to eMBB-2, leading to a higher throughput.

It must be pointed out that, within these slices, a certain gap is expected between the allocated PRBs and the average PRBs used, due to the temporal variability in V2X services demand. The resource allocation policy must ensure the expected reliability under traffic fluctuations, especially during peak demand periods. Therefore, a stabilization at a slightly negative value is expected.

The gap between PRBs allocated and average utilized in eMBB slices is illustrated in Fig. \ref{prbs_embb1} and Fig. \ref{prbs_embb2}, where the component $f_{\eta,4}$ of the slice eMBB-2 remains flat at zero throughout the entire training process. This effect is driven by the high throughput expectations of the service, which lead to the full utilization of the assigned PRBs in every proposed resource allocation during training. 
\begin{figure*}[h]
\centering
\subfloat[]{\includegraphics[width=7in]{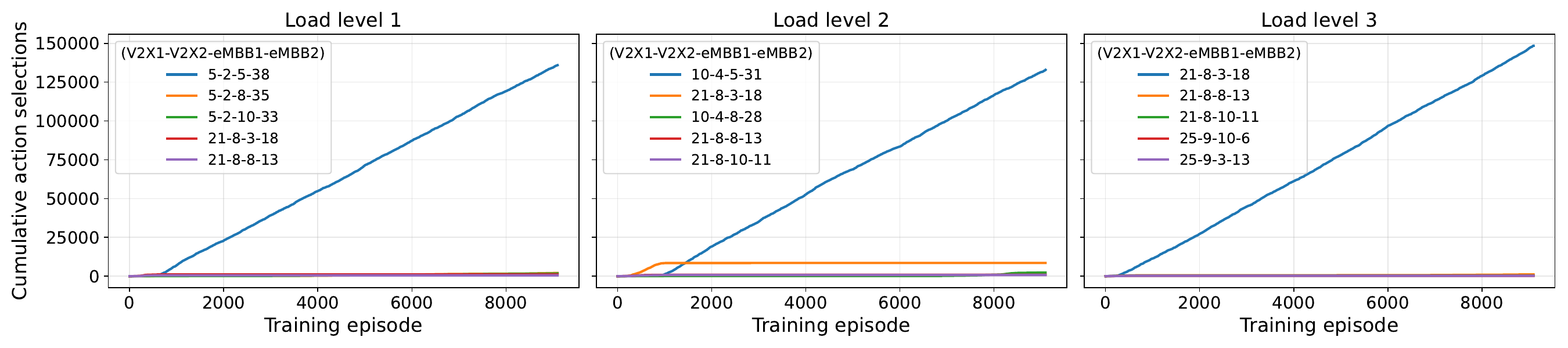}%
\label{acciones90}}

\subfloat[]{\includegraphics[width=7in]{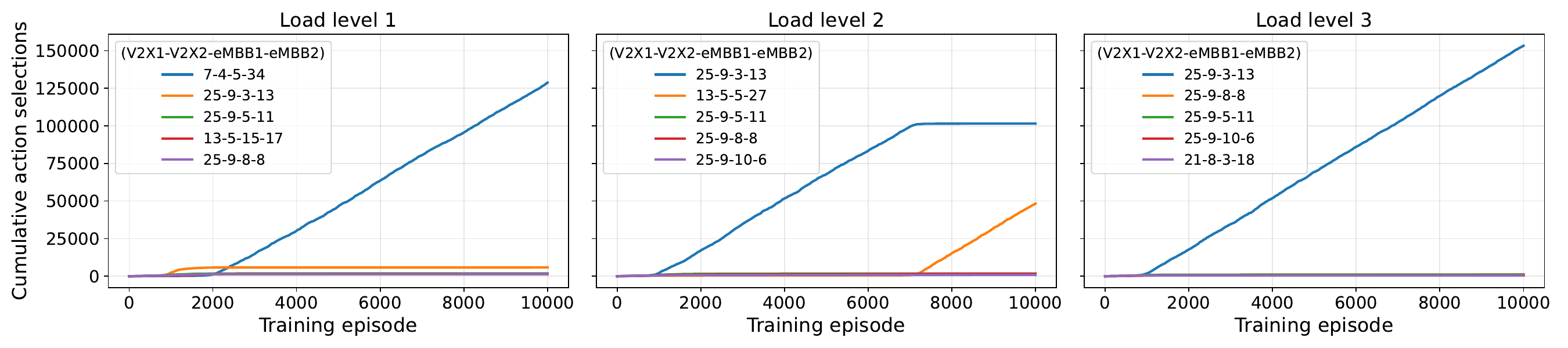}%
\label{acciones99}}
\hfil

\caption{Top 5 selected actions during training. (a) Cumulative action selection per episode for SLA 1 (b) Cumulative action selection per episode for SLA 2.}
\label{fig_distribuciones}
\end{figure*}
Conversely, the eMBB-1 service has lower throughput expectations. During the initial stages of training, the episode-averaged resource efficiency component $f_{\eta,3}$ reaches negative values due to exploration of different actions, some associated with slice over-provisioning. As training progresses, these values gradually increase and approach zero until stabilizing at a policy that ensures full utilization of the PRBs allocated to the eMBB-1 slice.

The learned resource allocation policy can also be observed through the recommended actions. Fig. \ref{acciones90} and Fig. \ref{acciones99} show the cumulative selection frequency of the five most frequently selected actions during training for each load level. As training progresses and the agent learns a policy aligned with the defined objectives, the actions that lead to higher rewards are most likely to be selected. Consequently, a dominant action emerges for each load level, indicating that the learned policy adapts the PRB distribution to the network load. 

However, a different behavior is observed for SLA 2 at load level 2. Initially, 25-9-3-13 emerges as the most frequently selected action for both load levels 2 and 3. Nevertheless, during this period, the V2X segments are over-provisioned at load level 2. Around episode 7000, the selection frequency of this action stops increasing, while a new allocation, the 13-5-5-27, gains presence and becomes the dominant action. This policy shift is consistent with the changes observed in Fig. \ref{tput_service} and Fig. \ref{prb_eff}, demonstrating that the agent refines its policy, moving away from replicating the allocation used for load level 3 and adapting the PRB allocation to the actual resource needs of each load level.

These results confirm that, in both SLA scenarios, the proposed solution converges to a policy that not only guarantees SLA compliance for all V2X services while seeking high resource efficiency, but also enables a resource allocation aligned with the specific requirements of eMBB services. 
 
\subsection{Model deployment analysis}

\begin{figure}[t]
    \centering
    \includegraphics[width=0.5\textwidth]{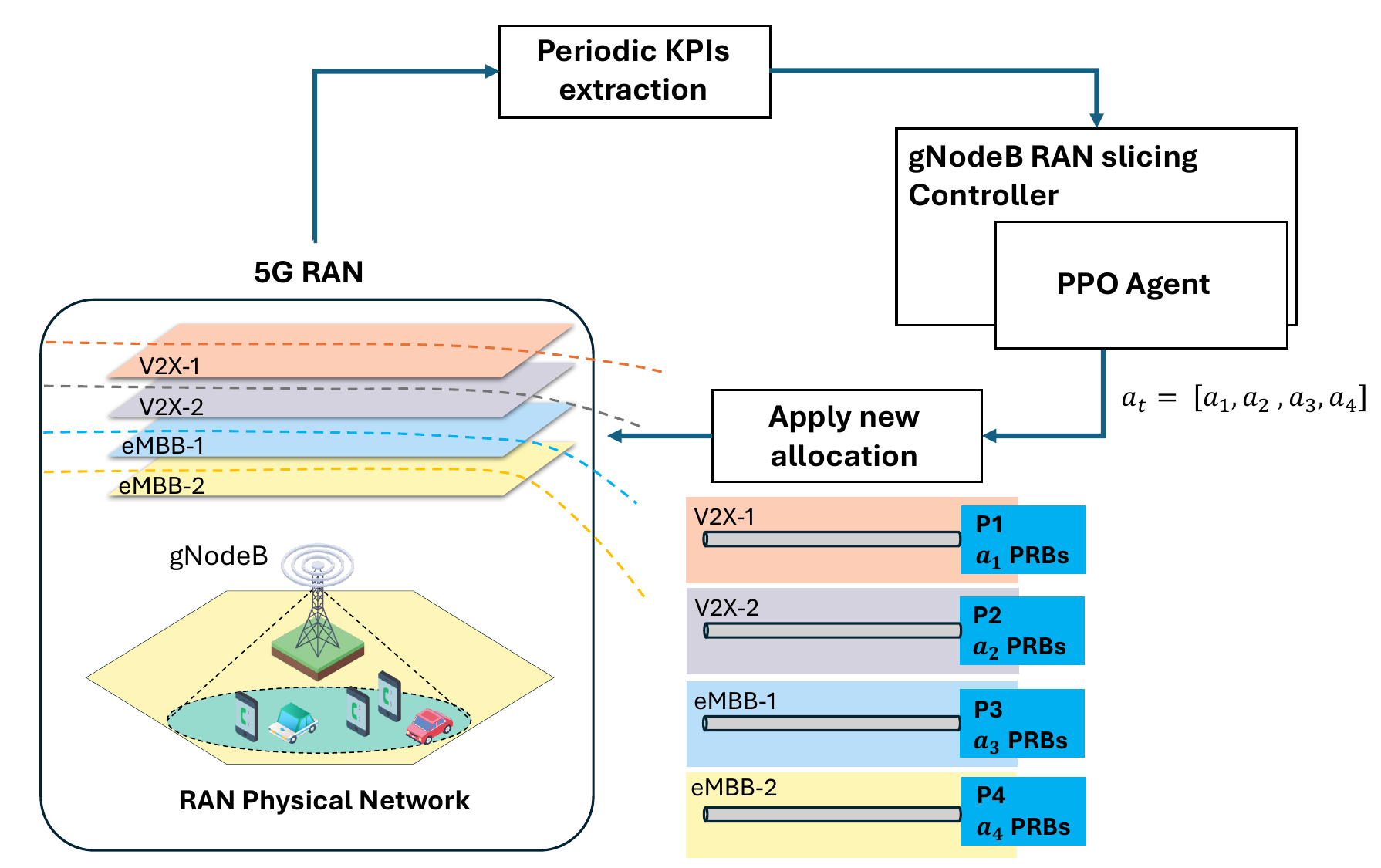}
    \caption{Deployment workflow of the proposed RAN slicing framework.}
    \label{fig_ML_cycle}
\end{figure}

Once the training process for a given SLA is completed, the model can be integrated into the network management framework. As described in Fig. \ref{fig_ML_cycle}, in a real deployment, the model will receive the network state periodically and determine the appropriate inter-slice resource allocation following the learned policy. If the new allocation differs from the configured one, the network will adapt the PRBs dedicated to each slice accordingly. 

To demonstrate its practical applicability, the trained models have been integrated into the simulation tool 
so that they can observe the network’s status based on the KPIs obtained over a given period of time and propose a new resource allocation, which is established as the new distribution of resources among the slices for the remainder of the simulation.

This interaction between the model and the network has been evaluated under various conditions. In this case, SLA 2, the most restrictive scenario in terms of requirements, was selected, along with the second load level, to demonstrate the model’s effectiveness in meeting the defined objectives when deployed in the network. To this end, two independent experiments are presented:

\begin{itemize}
    \item The first experiment demonstrates how, starting from a suboptimal distribution, the model is capable of recommending a new allocation that meets the SLA for both V2X services.

    \item The second experiment focuses on resource efficiency and demonstrates how, in situations where V2X slices are over-provisioned, the model recommends a new allocation that not only ensures continued compliance with the SLA but also leads to a more efficient use of resources.
\end{itemize}

\begin{figure}[t]

\subfloat[]{\includegraphics[width=3.5in]{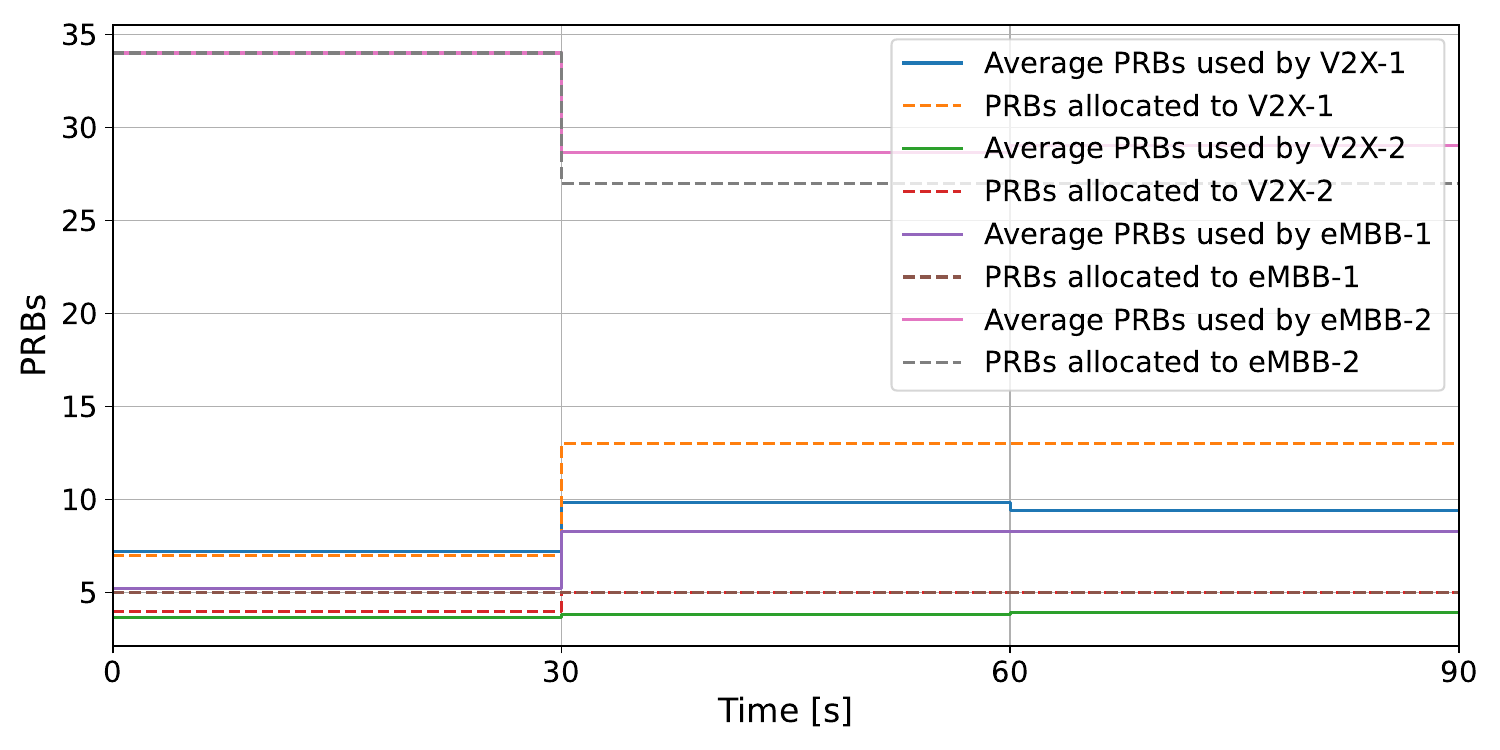}%
\label{prbs_slices}}

\subfloat[]{\includegraphics[width=3.5in]{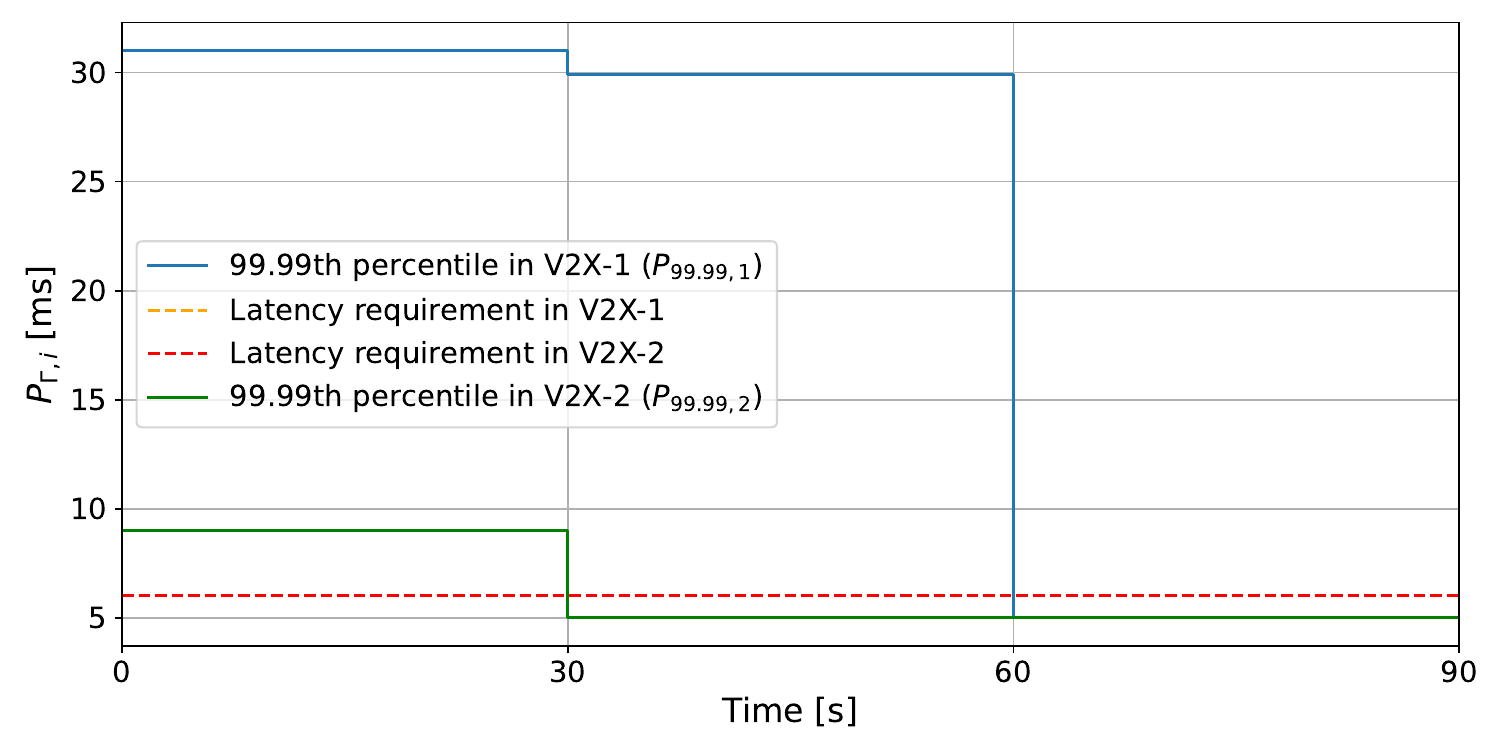}%
\label{p9999}}
\hfil

\subfloat[]{\includegraphics[width=3.5in]{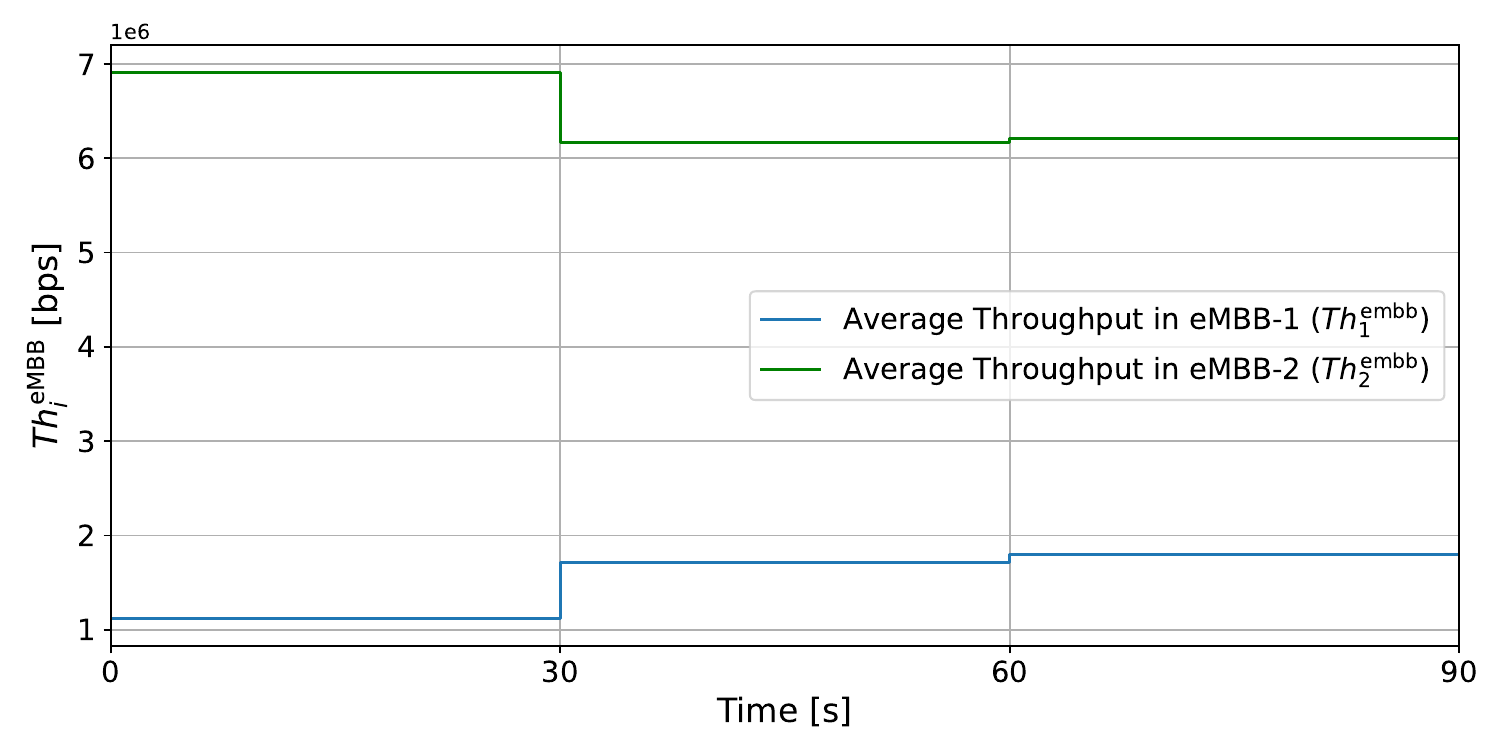}%
\label{tput_dep}}
\hfil
\caption{ Simulation results from the first experiment (a) Average PRB usage per slice (b) 99.99th percentile of radio delay in V2X services (c) Average user throughput in eMBB services.}
\label{fig_exp1}
\end{figure}

The results of the first experiment are presented in Fig. \ref{fig_exp1}. The conducted simulation comprises three $\mathcal{T}$ periods of 30 seconds each, a duration selected due to computational constraints. Nevertheless, the simulation was carried out under a traffic load high enough to ensure that the aggregate indicators are representative.  

In Fig. \ref{p9999}, the first period of 30 seconds, 99.84\% of the recorded delays for the V2X-1 service exceed 6ms, reaching a 99.99th percentile of 31 ms, which corresponds to the maximum time a packet remains in the system before being discarded for exceeding the PDB. In contrast, for V2X-2, 0.29\% of the delays exceed 6ms, and the 99.99th percentile is 9 ms. The model receives the KPIs from this period as input and proposes a new resource allocation, such that in the following two time periods, the number of PRBs dedicated to V2X slices increases, as shown in Fig. \ref{prbs_slices}. Specifically, the new configuration corresponds to $C=${\{13, 5, 5, 27\}}, the one observed in Fig. \ref{acciones99}.

This increment causes a slight drop in the 99.99th percentile during the second simulation period. This is a transition period, where packets at the beginning of the period still experience high delays carried over from the previous state, impacting the computation of metrics, especially in the V2X-1 service, which has a higher traffic intensity. However, in the third period, the situation stabilizes, delays decrease, and, as illustrated in Fig. \ref{p9999}, for both services the 99.99th percentile reaches 5 ms. This time, only 0.005\% of the recorded delays for the V2X-1 service exceed 6ms, while for the V2X-2 service, all recorded delays met the 6ms constraint.

For the eMBB-2 service, Fig. \ref{prbs_slices} shows that the model reduces the number of PRBs to 27, as additional resources are now reserved for the V2X slices. Consequently, the average throughput achieved by eMBB-2 users decreases slightly. In contrast, as observed in Fig. \ref{tput_dep} and Fig. \ref{prbs_slices}, the PRBs dedicated to eMBB-1 remain unchanged according to the learned policy (Fig. \ref{acciones99}), while its average throughput is higher than in the first period. This behavior is explained by the stochastic nature of the V2X traffic. As discussed in the previous section, the learned policy reserves sufficient resources to satisfy the configured SLA and, since the instantaneous arrival rate fluctuates over time, the reserved V2X resources are not fully utilized at every transmission interval, and they can be assigned to the remaining slices. 

\begin{figure}[t]

\subfloat[]{\includegraphics[width=3.5in]{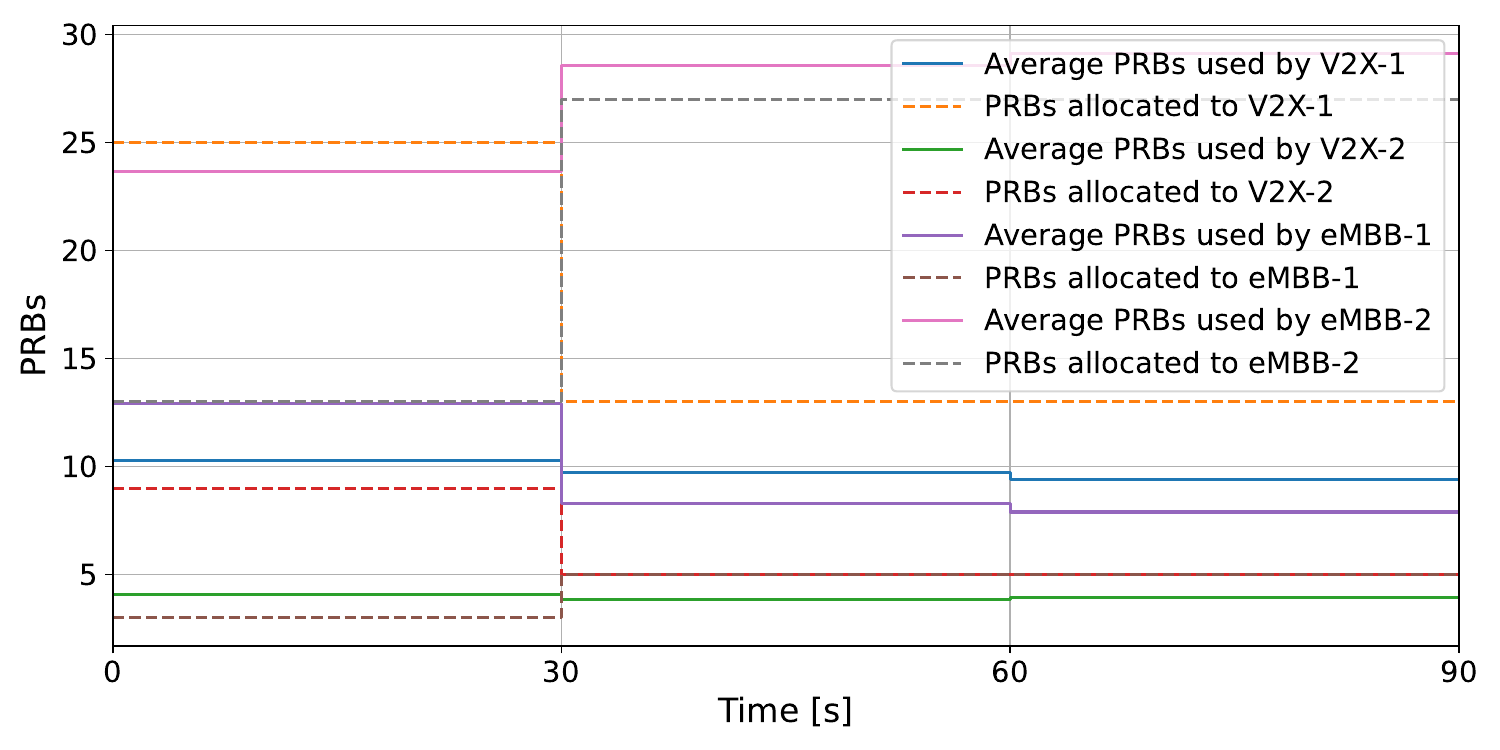}%
\label{prbs_overp}}
\hfil
\subfloat[]{\includegraphics[width=3.5in]{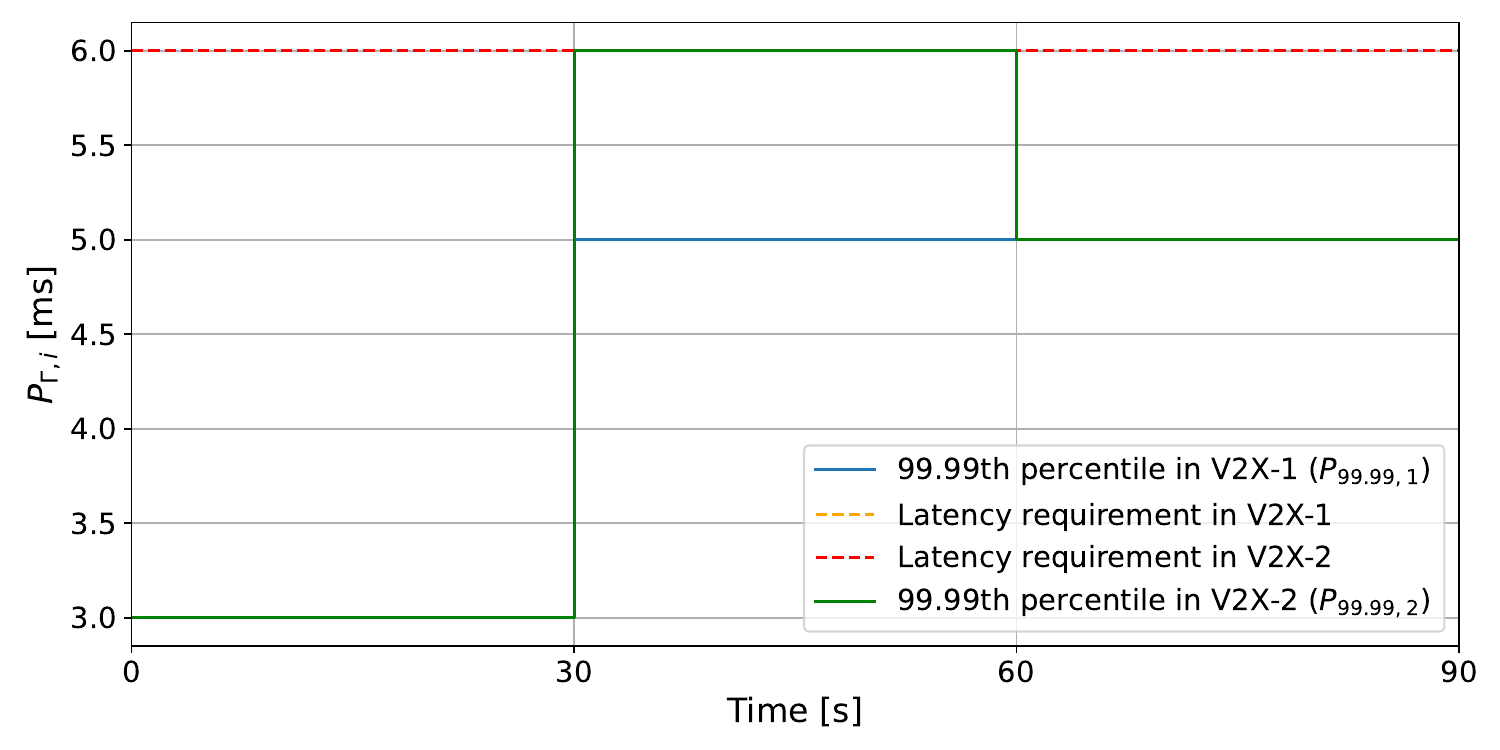}%
\label{p9999_overp}}

\subfloat[]{\includegraphics[width=3.5in]{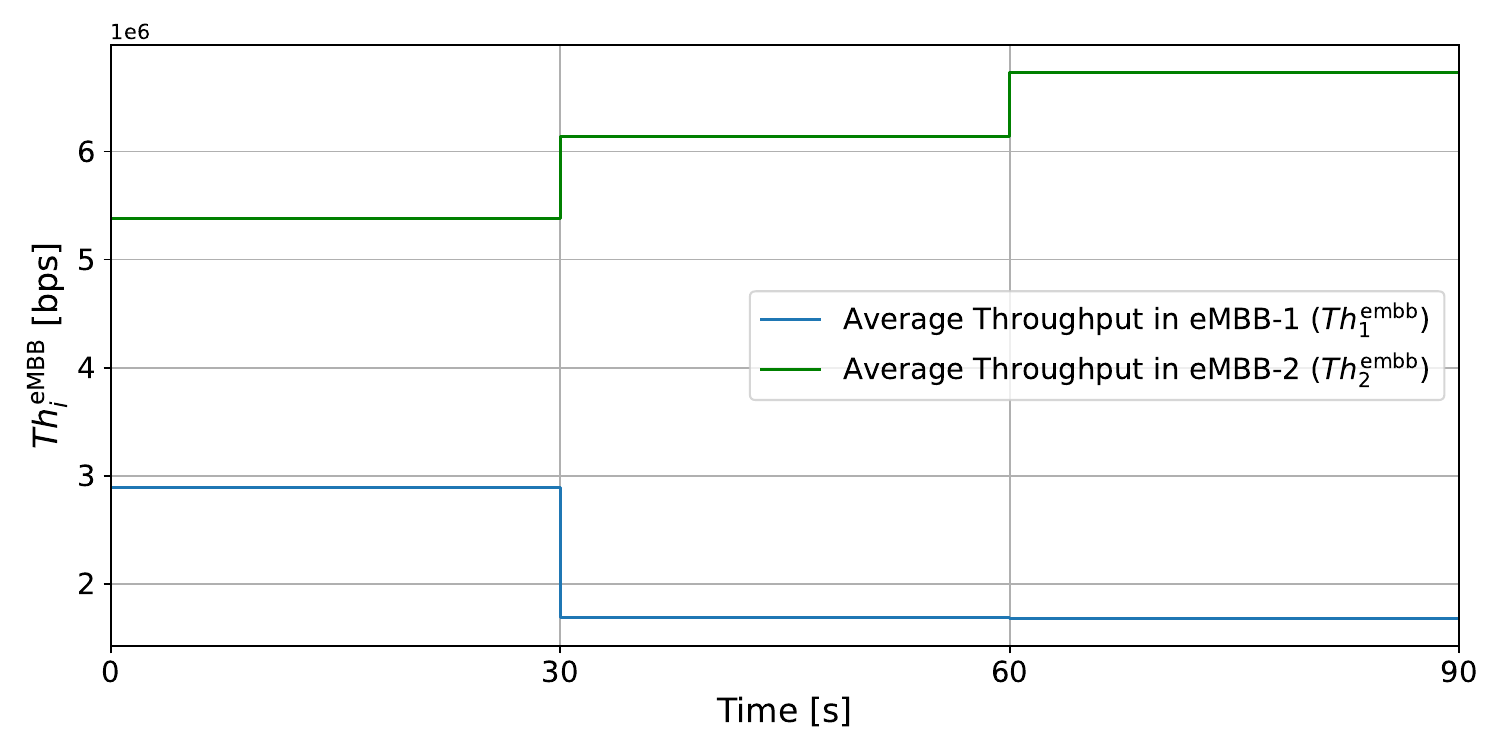}%
\label{tput_overp}}

\caption{ Simulation results from the second experiment (a) Average PRB usage per slice (b) 99.99th percentile of radio delay in V2X services (c) Average user throughput in eMBB services.}
\label{fig_overp}
\end{figure}

As observed in Fig. \ref{fig_overp}, the second experiment corresponds to another simulation with the same duration. In this case, during the first period, an excessive amount of resources is assigned to the V2X slices and, as illustrated in Fig. \ref{prbs_overp}, it differs markedly from the average utilization. In fact, the V2X-1 slice utilizes, on average, only around 41\% of its allocated resources, and V2X-2 reaches 45.5\% of utilization, while the eMBB slices achieve an average utilization higher than their assigned share. This behavior is a consequence of the adopted scheduling policy, described in Section \ref{sec:sys_model}, which allows these slices to compete for the PRBs left unused by the V2X slices. At the same time, in Fig. \ref{p9999_overp}, the 99.99th percentile of packet delay is notably below 6ms in this period. 

Once the model has analyzed the KPIs computed for the first period, a new PRB allocation is applied to the network. This time, the PRBs dedicated to the V2X slices are reduced, and with it the gap between assigned and average utilized PRBs, demonstrating that a more effective utilization has been achieved. With this new PRB allocation, the average resource utilization reaches 72.3\% for the V2X-1 slice and up to 78.96\% for V2X-2. Simultaneously, this new PRB allocation still allows meeting the 6ms constraint, as shown in Fig. \ref{p9999_overp}.

Regarding the eMBB slices, the allocated resources increase after applying the model's recommendation (Fig. \ref{prbs_overp}). Now, the competition for the resources freed up by the V2X slices decreases, as a large portion of them is reserved for eMBB-2 slice, which hosts the most bandwidth-intensive service. This also impacts resource utilization in the eMBB-1 slice, which now has fewer PRBs released by V2X to draw upon. 

The average user throughput achieved by the eMBB service in each slice is shown in Fig. \ref{tput_overp}. Here, a higher throughput is achieved in the eMBB-2 slice, with a clear improvement after the transition period. In contrast, the throughput achieved in the eMBB-1 slice decreases. Even though a larger portion of PRBs has been allocated to this slice, the resources occupied are lower, since the amount of PRBs freed up by the V2X slices is smaller than in the initial period.

Although the eMBB-1 throughput is far from the expected value, this new PRB allocation results in a better balance between the eMBB services. In the initial period, eMBB-1 benefited from the reuse of unused PRBs and reached its maximum throughput. With the new configuration, although eMBB-1 experiences a throughput reduction, both services achieve a more similar relative performance, while the SLA is still met in the V2X slices.

\section{Conclusions and future work}
In this paper, the challenge of supporting V2X applications in scenarios with heterogeneous services and performance requirements, including bandwidth-demanding ones, has been addressed. 

To meet the stringent requirements of V2X, a RAN slicing strategy based on DRL has been proposed to determine the optimal PRB distribution among slices while ensuring SLA compliance, efficient resource utilization, and minimized impact on eMBB performance. Furthermore, the model training relies on a dataset of precomputed simulations, avoiding direct interaction between the agent and the network and any negative impact on the user experience during training.

The results, obtained from a multi-slice scenario, have demonstrated that the algorithm converges to a policy that adapts the PRB allocation according to the network demands. For every considered load level, compliance with the V2X SLAs is ensured, and the resource efficiency of each partition is improved. Consequently, the final performance is prevented from being influenced by factors such as competition for the remaining resources.

Finally, the trained agent has been deployed in the simulation tool, where it interacted with the network under the most restrictive SLA. The observed results demonstrate that the learned policy responds effectively under both SLA violation and over-provisioning scenarios. 

Therefore, the presented methodology provides a flexible strategy for dynamic PRB allocation that can be extended to different SLA and throughput targets, as well as to network load conditions beyond those studied in this work, through appropriate training. Thus, an efficient solution for supporting reliable and latency-critical services such as V2X in heterogeneous multi-service scenarios is provided through optimal resource management in RAN slicing, while efficient utilization of the radio resources allocated to the slices is simultaneously promoted. 

As future work, the proposed solution will be adapted and validated in a real network deployment to assess its feasibility and performance under realistic operating conditions and to evaluate the practical challenges beyond simulation-based environments.


\begin{IEEEbiographynophoto}
{M. Martínez}
received the B.S degree in Telecommunications Technology
Engineering from the University of Malaga in 2023 and the
Master’s degree in Telecommunication Engineering at the International University of La Rioja. She is currently pursuing her Ph.D. degree in Telecommunications Engineering at the University of Málaga.  Her research interests include the application of AI techniques for the intelligent management of telecommunication networks.

\end{IEEEbiographynophoto}

\begin{IEEEbiographynophoto}{I. de-la-Bandera}
received the MSc and PhD degrees in telecommunications engineering from the University of Málaga. She joined the Communications Engineering Department of the
University of Málaga, Spain, in 2010. Since then, she has participated in many projects, national and international, concerning radio resource management in mobile networks. She has collaborated with major
mobile operators and vendors.

\end{IEEEbiographynophoto}

\begin{IEEEbiographynophoto}{D. E. García} received the B.Sc. in Telecommunications Technologies Engineering, and a M.Sc. degree in Telecommunication Engineering from the University of Malaga in 2025 and 2026, respectively. He is currently pursuing his Ph.D. degree in Telecommunications Engineering at the University of Malaga. His main research interests include AI applications for cellular and Non-Terrestrial Networks (NTN).

\end{IEEEbiographynophoto}

\begin{IEEEbiographynophoto}{P. Vera}
received the B.Sc. degree in Electronics, Robotics and Mechatronics engineering and a M.Sc. degree in Mechatronics engineering from the University of Malaga in 2021 and 2022, respectively and he is currently pursuing his Ph.D. degree in telecommunications engineering at the University of Málaga. His main research focus in mobile communications, avionics networks and cloud robotics.

\end{IEEEbiographynophoto}

\begin{IEEEbiographynophoto}{S. Fortes}
holds a M.Sc. (2010) and a Ph.D. (2017) in Telecommunication Engineering from the University of Málaga (UMA). He began his career being part of main european space agencies (DLR, CNES, ESA) and Avanti Communications plc, where he participated in various research and consultant activities on broadband and aeronautical satellite communications. In 2012, he joined the University of M\'alaga, where he is currently Associate Professor. Here, he has led more than 30 projects and contracts with key industry partners, mobile communications operators and vendors. His research focuses on the application of advanced algorithms and AI to complex systems, particularly cellular and satellite networks and satellite systems, and applications such as smart cities, cloud robotics, and healthcare

\end{IEEEbiographynophoto}

\begin{IEEEbiographynophoto}{M. L. Luque}
 is an AI-RAN Technical Lead within Ericsson’s Cognitive Network Solutions organization, focusing on AI research and prototyping for advanced network solutions. Her background spans standardization, patent development, and network optimization. She holds M.Sc. degrees from the University of Málaga and Aalborg University, graduating with highest honors from both. She is co-inventor of more than 60 patents.

\end{IEEEbiographynophoto}

\begin{IEEEbiographynophoto}{A. Mendo}
 received the M.S. degree in Telecommunication Engineering with highest honors from the University of Málaga, Málaga, Spain, in 2004. Since 2004, he has been a Researcher at Optimi Corporation and joined the Ericsson Group in 2010. He has been involved in several research projects in mobile communications, with a focus on network optimization and artificial intelligence. He has authored several papers in international conferences and journals and is a co-inventor of more than 30 patents held by Ericsson. His current research interests include self-organizing networks, network design and optimization, and artificial intelligence.

\end{IEEEbiographynophoto}

\begin{IEEEbiographynophoto}{J. Ramiro}
is heading up the research and Innovation team within the Cognitive Network Solutions Area in Ericsson’s Business Area Cloud and Software Services, where he leads R and D activities with mid and long-term horizon, involving exploration of new technologies, forward-looking concepts and disruptive network optimization methodologies. Juan holds a Telecom Engineering degree from Malaga University, a PhD in Electrical and Electronic Engineering from Aalborg University, an Executive MBA from San Telmo Business School and an Executive Degree in Big Data and Business Analytics from EOI.
\end{IEEEbiographynophoto}

\begin{IEEEbiographynophoto}{R. Barco}
received the M.Sc. and Ph.D. degrees in Telecommunication Engineering from the University of Málaga Spain, where she is currently a Full Professor. She has worked at Telefónica, Spain, and
at the European Space Agency, and she participated
in a Mobile Communication Systems Competence
Center, jointly created by Nokia and the UMA. She
has published more than 100 scientific papers, filed
several patents, and has led projects with major
companies.
\end{IEEEbiographynophoto}

\vfill

\end{document}